# Water-resistant hybrid perovskite solar cell - drop triboelectric energy harvester

Fernando Núñez-Gálvez,[a,b] Xabier García-Casas,[a] Lidia Contreras Bernal,[a,c] Alejandro Descalzo,[a] José Manuel Obrero-Pérez,[a] Javier Castillo-Seoane,[a] Antonio Ginés,[d] Gildas Leger,[d] Juan Carlos Sanchez-Lopez,[e] Juan Pedro Espinós,[a] Ángel Barranco,[a] Ana Borrás,[a]* Juan Ramón Sánchez-Valencia,[a]* Carmen López-Santos[a,b]*

a) Nanotechnology on Surfaces and Plasma Laboratory, Consejo Superior de Investigaciones Científicas (CSIC), Materials Science Institute of Seville (CSIC-US). c/ Américo Vespucio 49, 41092, Seville (Spain).
b) Departamento de Física Aplicada I, Universidad de Sevilla, c/ Virgen de Africa 7, 41011, Seville (Spain).
c) Dpto. Química Física. Facultad de Química. Universidad de Sevilla. C/Profesor García González 41012, Seville (Spain).
d) Instituto de Microelectrónica de Sevilla (IMSE-CNM), Universidad de Sevilla – CSIC, Avda. Americo Vespucio 28, 41092, Seville (Spain).
e) Materials Science Institute of Seville (CSIC-US). c/ Américo Vespucio 49, 41092, Seville (Spain).

**Corresponding authors:** Ana Borras, anaisabel.borras@icmse.csic.es; Juan R. Sanchez-Valencia, jrsanchez@icmse.csic.es; Carmen López-Santos, mclopez@icmse.csic.es.

## Abstract

Hybrid energy-harvesting systems that combine perovskite solar cells (PSCs) with drop-driven triboelectric nanogenerators (D-TENGs) offer a compelling solution for continuous power generation under diverse weather conditions. Yet, the inherent vulnerability of halide perovskites to moisture and environmental stressors remains a critical barrier to their widespread deployment. To overcome this bottleneck, we introduce plasma-deposited fluorinated polymers ($CF_x$) films as multifunctional encapsulation layers that simultaneously provide water resistance, triboelectric functionality, and optical transparency (>90%). Plasma deposition enables conformal, room temperature, and solvent-free coating of complex surfaces, ensuring uniform protection without compromising photovoltaic performance. After encapsulation of PSCs with $CF_x$ films, power conversion efficiency remained virtually unchanged, and champion cells preserved a PCE of 17.9%. More importantly, the devices exhibited high environmental stability, retaining over 50% of their initial PCE for 10 days under high humidity and temperature. Furthermore, $CF_x$ layers enabled Spiro-OMeTAD compatibility with commercial UV-curable resins, leading to a thin-film hybrid PSC/D-TENG device capable of simultaneous solar and rain energy harvesting. This device maintained 80% of its initial performance after 300 hours of continuous illumination under humid conditions and demonstrated stability under continuous dripping and illumination for more than 5 hours. We demonstrated that optimizing the chemical composition of $CF_x$ layers significantly enhances their triboelectric performance. In standalone operation, the optimized $CF_x$-based D-TENG, enriched with 36.4% of ($CF_2$ + $CF_3$) functional species, delivered open-circuit voltage peaks up to 110 V and a maximum power density of ~4 mW/cm$^2$ under rainwater droplets, while retaining over 85% of its initial output after more than 17,000 droplet impacts. As a proof of concept, using the same $CF_x$ layer for both encapsulation and triboelectric functionality, the hybrid PSC/D-TENG device achieved short-circuit current densities of 11.6 mA/cm$^2$ under 0.5 sun illumination and peak voltages of 12 V per raindrop, enabling simultaneous solar and rain energy harvesting. A self-charging prototype powered LED arrays via a custom boost converter, demonstrating practical multisource energy harvesting for low-power electronics.

## 1. Introduction

Multifunctional and hybrid multisource energy converters and harvesters have evolved as key-enabling technologies for powering Industry 4.0, addressing the increasing demand for renewable energy. Integrating solar cells with various piezoelectric and triboelectric nanogenerators provides a reliable path for efficiently harvesting environmental energy from outdoor and indoor light sources, body movements, and vibrations.[1–3] These integrated solutions have already been used on a small scale for self-powered sensors and wearable devices. However, a “game changer” has recently emerged: combining solar photovoltaic panels with triboelectric systems to convert the kinetic energy of raindrop impacts.[4,5] This innovative approach will pave the path toward synergistic solar and rain panels, opening up new possibilities for sustainable energy harvesting under all environmental conditions.

Numerous triboelectric nanogenerator alternatives have been explored, varying in device architectures, triboelectric materials combinations and functionalization, and energy management systems. However, in the photovoltaic counterpart, mostly silicon solar cells have been exploited in these hybrid systems, with few attempts at implementing Q-dot and organic-based solutions.[4,6–8] One of the most prominent technologies nowadays, the hybrid halide perovskite solar cells (PSCs), has been largely overlooked. This situation is significant considering their potential for price competitiveness,[9] compatibility with flexible configuration,[10,11] and outstanding high efficiency, certified as 27 % in 2025 for a single junction.[12] Moreover, silicon/perovskite tandem solar cells have demonstrated record efficiencies of 34.9%, surpassing the theoretical limits of single-junction devices and making them a prime candidate for industrial-scale deployment of perovskite solar cells.[12] However, the Achilles heel of PSCs, i.e., the low long-term stability under actual environmental working conditions, has limited their hybridization with other energy harvesting systems. Concretely, PSCs present low stability in the presence of oxygen, UV light, temperature gradients, and, especially, moisture.[13–18] This latter sensibility to humidity inherently hinders their combination with drop triboelectric nanogenerators and rain panels. The environmental instability of perovskite solar cells is related to the ionic nature of halide perovskites, the presence of structural defects, and the hygroscopic properties of both the perovskite layer and the commonly used charge transport materials. Additional degradation pathways include sensitivity to oxygen, UV light, and thermal stress, which collectively accelerate material decomposition and interface deterioration.[16] Solutions have been proposed in the form of additives,[9,15,17] passivation elements,[19,20] encapsulants,[21–23] superhydrophobic silica particles,[24] or even alternative chemical stoichiometries,[25–27] and microstructures,[13,28,29] which preserve the performance of the absorber material while ensuring its stability and reproducibility under practical use conditions.

Encapsulation can cause an initial efficiency drop in perovskite solar cells due to optical losses and interfacial changes. Methods using epoxy resin and ethylene-vinyl acetate (EVA) often result in immediate efficiency reductions, primarily due to light absorption and chemical interactions. In contrast, polyurethane (PU), polyisobutylene (PIB), and glass-glass encapsulation have a lower impact, preserving optical transmission and device integrity.[30,31] Herein, we propose a Teflon-like ($CF_x$) plasma polymer thin encapsulant with a double functionality. On the one hand, this hydrophobic thin film enhances the PSC stability under highly humid conditions. On the other hand, the $CF_x$ works as an antireflective triboelectric layer that makes the cell compatible with operation as D-TENG. Both results have significant implications for water-resistant PSCs and the hybridization of advanced energy harvesters. Selecting robust and efficient encapsulants is not straightforward.[32] The encapsulation is expected to be the ultimate step in the PSC assembly procedure. However, due to synthetic constraints, a key limitation of standard encapsulants is their incompatibility with commonly used charge transport materials such as Spiro-OMeTAD. In particular, the Ossila epoxy resin (E132) used in this work has been found to produce a loss on the efficiency of PSCs upon encapsulation, due to its poor chemical compatibility with methoxy-containing compounds such as Spiro-OMeTAD.[33] Currently, dry methodologies devoted to the fabrication of encapsulating thin films at room temperature are exploited instead of standardized solution procedures for the encapsulation with highly cross-linked, densified layers,[34] water molecule barrier oxide membranes,[35] metal oxide capping layers,[33] and ultra-thin hydrophobic polymeric films.[36–38] Our recent results reported adamantane plasma polymers as a reliable interface and encapsulant fabricated in a downstream microwave plasma reactor.[19,22]

Water repellence is one of the primary challenges that the encapsulant material must guarantee. Fluorine-based compounds are known for their unique water-repellent behavior. Therefore, fluorine-based passivation layers, [39–43] electron,[44–46] and hole transport layers,[47,48] as well as other strategies based on electrodes and coatings[26,49] have been explored as moisture-protecting solutions. Notably, fluorinated polymers are ubiquitous in developing triboelectric nanogenerators, including liquid-solid nanogenerators as drop-TENGs. [50–54]

D-TENGs have rapidly evolved since the first practical demonstration by Wang *et al.* in 2014 for a single electrode configuration.[55] Thus, power densities as high as 50 $W/m^2$ [56] and 200 $W/m^2$ [57] have been reported exploiting a dual, top–bottom switch electrode architecture and D-TENG arrays.[58] Most triboelectric rain panels rely on bulk polymer triboelectric layers,[52,53,59] with very few examples employing thin film harvesting layers.[60] Thus, in 2022, Sun *et al*.[61] reported the first example of a hybrid halide perovskite solar cell working with a D-TENG due to the previous works consider solid-solid contact in PSC/TENGs combination.[62–68] In that example, $MoO_3$ was employed as a high permittivity and wide bandgap layer between the PSC and a polymeric FEP triboelectric generator. The $MoO_3$ worked as an effective electron-blocking layer (EBL) between the triboelectric polymer and the electrode, reducing the recombination rate at the interface. Raindrop output power reached values up to 0.68 mW. Still, the issue of the PSC stability was overlooked, and the energy harvesters were implemented as relatively independent systems integrating additional layers, including the bulky triboelectric layer (thickness over 1.2 µm). Yang *et al*.[69] enhanced hybridization by combining a D-TENG with a ferroelectric ($BaTiO_3$) cell sharing bottom electrodes. Nevertheless, the $BaTiO_3$ cell, working in the UV range, presented limited power conversion efficiency. Recently, Lee *et al*.[70] reported the combination of a PSC with a fluorocarbon plasma polymer and surface nanostructuration of polyimide. In this article, the authors showed an enhancement in the PSC photocurrent after implementing the triboelectric system in a flexible configuration. However, in that article, the hybrid architecture is only tested for solid-solid triboelectrification, and the performance for drop energy harvesting is not considered.

For our demonstration of a hybrid PSC / D-TENG thin film system, we have implemented the structure proposed by Wang *et al*.,[56] taking advantage of the FTO electrode as a transparent electrode for the PSC and bottom electrode of the D-TENG and exploiting the $CF_x$ thin film as a triboelectric harvesting layer for the D-TENG. This layout is fully compatible with direct and inverse PSC architectures and with the extension of the D-TENG as an array or to different top electrode configurations to enhance power density [71] and, as far as we know, is the first demonstration of a PSC / D-TENG system relying on an in-built triboelectric thin film layer. Thus, the $CF_x$ layer with a thickness of up to 400 nm works as environmental protection on top of the PSC structure. This layer is prepared without solvents, at room temperature, and by a low-energy plasma-assisted vacuum technology.[72,73] This procedure is fully compatible and respectful with the PSC performance, contrary to what one might commonly think in the state of the art of PSC encapsulants, that low-temperature plasma discharge could be detrimental to the perovskite absorber or generate pinholes. Moreover, we demonstrate how this thin film encapsulant acts as an intermediary protector, ensuring the compatibility of commercial epoxy encapsulants with Spiro-OMeTAD-containing PSCs. This combination leads to remarkable cell stability, making the cell compatible with water immersion for 15 minutes. The article presents the optimization of the plasma polymer encapsulant for methylammonium-formamidinium-based (RbCsMAFA) solar cells, the characteristic triboelectric performance for chemically optimized $CF_x$ layers, and, finally, a proof-of-concept for the hybridization of D-TENG and PSC based on the twofold role of the plasma polymer, i.e., encapsulant and triboelectric layer.

## 2. Results and discussion

### 2.1. Fluorinated thin film encapsulants by Plasma Enhanced Chemical Vapour Deposition (PECVD)

Fluorinated coatings ($CF_x$) of different thicknesses were deposited on complete PSCs and flat reference substrates. **Figure 1** a) shows the UV-Vis-NIR direct transmittance spectra of the $CF_x$ coatings deposited on fused silica substrates. The thin films are highly transparent in the 400-2500 nm wavelength range, reaching high transmittance values (ca. 93-94 %). Apart from the layer of 30 nm, the transmittance spectra surpass that of the substrate (see zoom-in), indicating that the $CF_x$ coatings present a lower refractive index than the substrate (in these measurements is fused silica, but in the final device, it will be glass, both with n(@525nm)~1.46-1.50). These results are confirmed by variable-angle ellipsometry, as shown in **Figure S1** in the Supporting Information section.

This confers the system with antireflective properties, an appealing feature of the D-TENG triboelectric coating that provides a direct implementation with the solar cell.[70] However, the 30 nm thick $CF_x$ layer exhibits reduced transmittance, likely due to incomplete substrate coverage and increased light scattering caused by surface roughness. Thus, when the film is very thin, the coverage may not be uniform, resulting in nanometric-scale features that act as scattering centers. Light scattering is strongly dependent on the feature size relative to the wavelength of light, and in this case, the dimensions of the incomplete coverage are in the range of visible light, leading to enhanced scattering and consequently reduced transmittance. The $CF_x$ thickness also affects the absorption bandgap, i.e., the UV transparency below 400 nm wavelength, which increases for smaller thicknesses. Blocking far UV (UVB and UVC) is highly advantageous, as it is well known to cause significant degradation of the PSC, especially in those using metal oxide transport layers.[14]

Figure 1 b) gathers the wetting characterization for the $CF_x$ encapsulants deposited on the PSC following the layer-by-layer structures presented in panels c) and d). Water Contact Angles (WCA) of around 110° are obtained, independent of the $CF_x$ thickness, which is practically twice the PSC reference value and in line with the inherent hydrophobicity of fluorinated polymers.[73] Moreover, the contact area under the 2 µl water droplet deposited on the uncoated PSC surface is immediately degraded (turning to a yellowish color linked to the lead iodide appearing due to the perovskite hydrolysis). Although this hydrophobicity facilitates droplet mobility and reduces contaminant adhesion (which will be beneficial for the objectives of this work), achieving true self-cleaning behavior typically requires hierarchical roughness to minimize contact angle hysteresis and enable particle removal. Introducing such texturing would compromise optical transparency, which is critical for our multi-source harvester design. Therefore, the $CF_x$ layers in this work provide partial functionality (low adhesion and good droplet sliding) without affecting light transmittance, thereby representing a practical balance between surface properties and device performance.

In contrast, the $CF_x$-encapsulated PSCs remain chemically stable for thicknesses above 100 nm over time. Figure 1 e) shows the characteristic X-Ray Photoelectron Spectroscopy (XPS) curves corresponding to the C 1s spectrum of the surface of a 100 nm $CF_x$ thin film on top of the whole PSC system. Functional fluorinated groups are identified as the contribution of various stoichiometric fluorocarbon groups (-$CF_3$, -$CF_2$, -CF, and -C-CF at 290.8 eV, 288.6 eV, 287.5 eV, and 286.4 eV of binding energy, respectively) and a minor presence of C-C/C-H groups at 284.5 eV.[73] Note that, unlike previously reported Teflon-like synthesis on Spiro-OMeTAD films (without LiTFSI and 4-tert-butylpyridine dopant) performed by a $CF_4$ microwave plasma by Wu *et al.*,[74] the $CF_x$ developed herein is richer in other fluorinated stoichiometries, with a much smaller amount of oxygen at the surface (around 2%), as demonstrated in the XPS survey and the surface elemental quantification in **Figure S2.** It is worth stressing that, contrary to the reported etching by plasma discharges with fluorinated precursors (even at low bias voltages),[75,76] our $CF_x$ growth is fully compatible with the deposition on PSCs. Thus, Figure 1 c) right shows a 100 nm $CF_x$ film on top of the PSC, depicting a uniform, pinhole-free, and continuous cross-section and surface along the entire sample area. Below the $CF_x$ coating, the PSC sandwich stack (FTO, c-$TiO_2$ (*compact*), m-$TiO_2$ (*mesoporous*), PVK (*perovskite*), Spiro-OMeTAD, and Au) schematized in Figure 1 d) remains intact (see comparison with the reference at the left side of the panel c). These results confirm that the proposed fluorinated plasma-assisted CVD process carried out at RT, mild vacuum, and low power conditions does not affect the chemical composition of the metal electrode and Spiro-OMeTAD film.

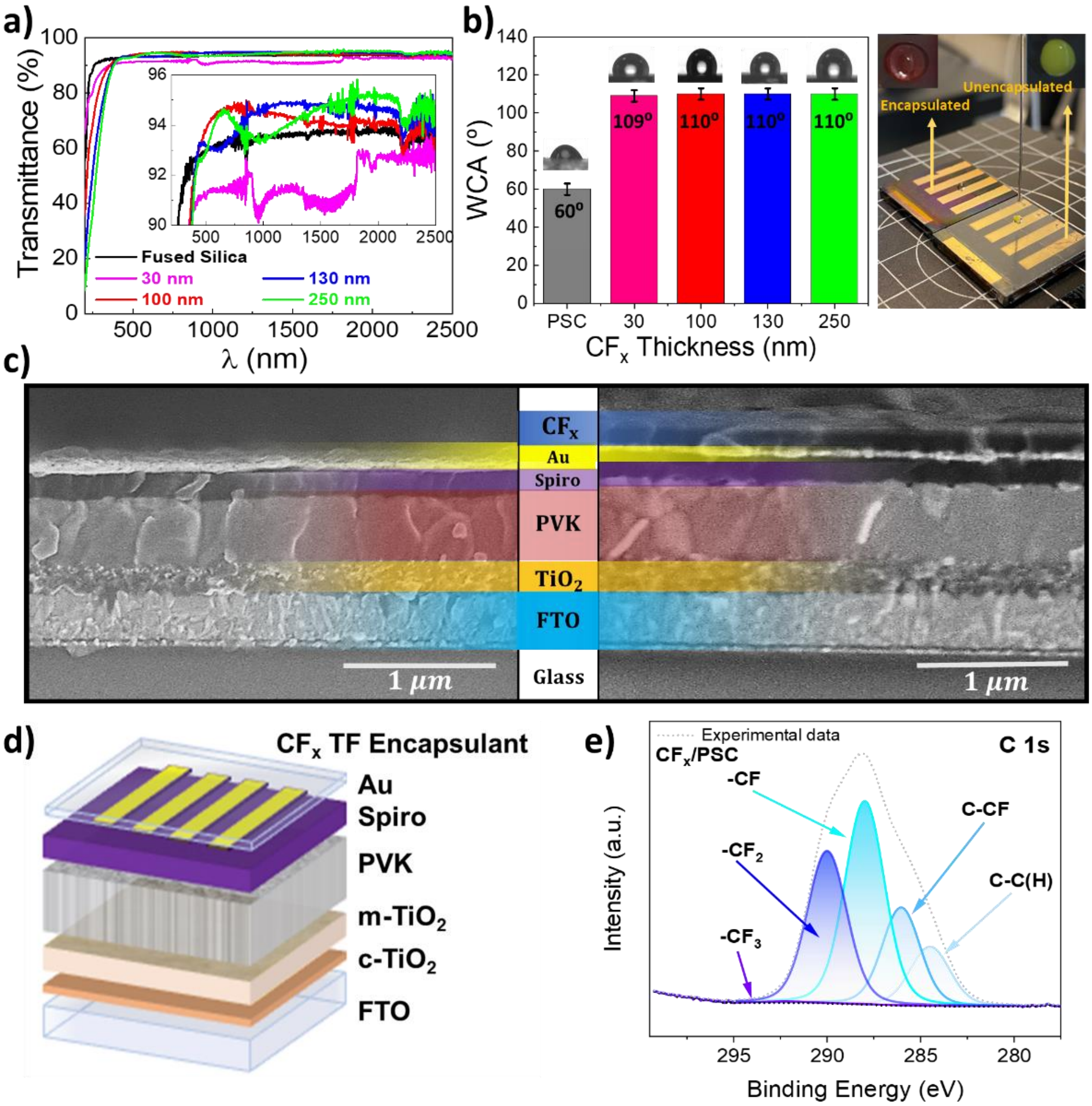

**Figure 1. Transparency, wettability, microstructure, and chemical composition of the $CF_x$ films encapsulants for PSC.** a) Transmittance spectra of $CF_x$ coatings as a function of the thickness. The zoom-in shows the higher transmittance of the coatings with respect to the fused silica substrate. b) Water contact angle for different $CF_x$ thicknesses and a photograph of a water droplet on a $CF_x$ encapsulated and an unencapsulated PSC. The inset shows the top view of the droplets 10 seconds after dropping. c) SEM micrographs corresponding to the cross-section views of reference (left) and $CF_x$ encapsulated PSCs (right). d) Scheme of the architecture of $CF_x$ encapsulated PSC. e) High-resolution XPS binding energy spectrum corresponding to the C1s region indicating the fluorine-based functional groups of a 100 nm $CF_x$/PSC surface.

In addition, the $CF_x$ layers exhibit remarkable mechanical properties, as shown in **Figure S3**. Nanoindentation tests revealed a Young's modulus of 3.7 $\pm$ 0.2 GPa and a hardness of 0.23 $\pm$ 0.01 GPa, which corresponds to the typical behavior of a relatively soft polymer but with greater stiffness than common fluorinated polymers such as PTFE. This enhanced resistance to plastic deformation is attributed to the plasma polymerization process, which promotes the formation of highly crosslinked polymeric networks during fabrication. Such mechanical robustness is expected to be highly beneficial for improving both the encapsulation performance and the triboelectric properties of the resulting films as we will present below.

## 2.2. Photovoltaic performance of the $CF_x$-encapsulated PSCs

**Figure 2** presents the compatibility of the $CF_x$ encapsulation with the solar cell device operation. Thus, current density vs voltage characteristics measured under 1 sun illumination of the champion devices before (dotted lines) and after encapsulation with $CF_x$ (solid lines) for the four different thicknesses studied (a) depict similar trends. The statistical analysis of variances using ANOVA [77] for more than 40 cells for the open-circuit voltage ($V_{OC}$) (b), short-circuit current density ($J_{SC}$) (c), Fill Factor (FF) (d), and Power Conversion Efficiency (PCE) (e), confirms the preservation of photovoltaic parameters after coating with $CF_x$. It is worth highlighting that this statistical analysis (Figure 2 b-e) was conducted on PSCs encapsulated with $CF_x$ for devices synthesized in different batches. Each batch of PSCs was measured both before encapsulation (indicated by grey boxes and lines) and after encapsulation (indicated by coloured boxes and lines). As a result, the average values and data dispersion vary for each $CF_x$ thickness. Therefore, the emphasis should be on the changes in photovoltaic parameters within the same batch rather than comparisons across batches.

For this reason, the statistical dispersion before and after $CF_x$ encapsulation has a close relation in all the cases, i.e., the reference batches (grey bars) with a higher/lower dispersion maintain this high/low distribution after encapsulation. Despite this variety, inherent to the synthesis by solvent phase of the PSCs, the $CF_x$ encapsulation has not substantially affected the photovoltaic characteristics of the devices, apart from a slight decrease of the short-circuit current density and open-circuit voltage values, but preserving in most of the cases the Fill Factor. The $V_{OC}$ slightly decreases after encapsulation for all the thicknesses except for the thinnest (30 nm). It is worth stressing that this feature is likely related to a reduced electrical connectivity, attributed to the CFx interlayer deposited over the Au electrode, a consequence of not applying masking during encapsulation. This choice was intentional, aiming to achieve a fully integrated barrier layer, as proper edge sealing is considered essential for durable perovskite encapsulation.[23,78] The conformal PECVD-deposited $CF_x$ coating enhances hermeticity by protecting the edges and lateral interfaces, but it also increases contact resistance at the Au/pin-electrode interface. This impact becomes more pronounced with thicker $CF_x$ films, ultimately affecting the overall photovoltaic performance. Moreover, increased $CF_x$ film thickness may induce mechanical stress within the multilayer stack due to a non-optimal interfacial adhesion between the $CF_x$ layer and underlying materials. This mechanical mismatch can promote the formation of microcracks, which may compromise the encapsulation integrity and reduce the long-term stability of the device.

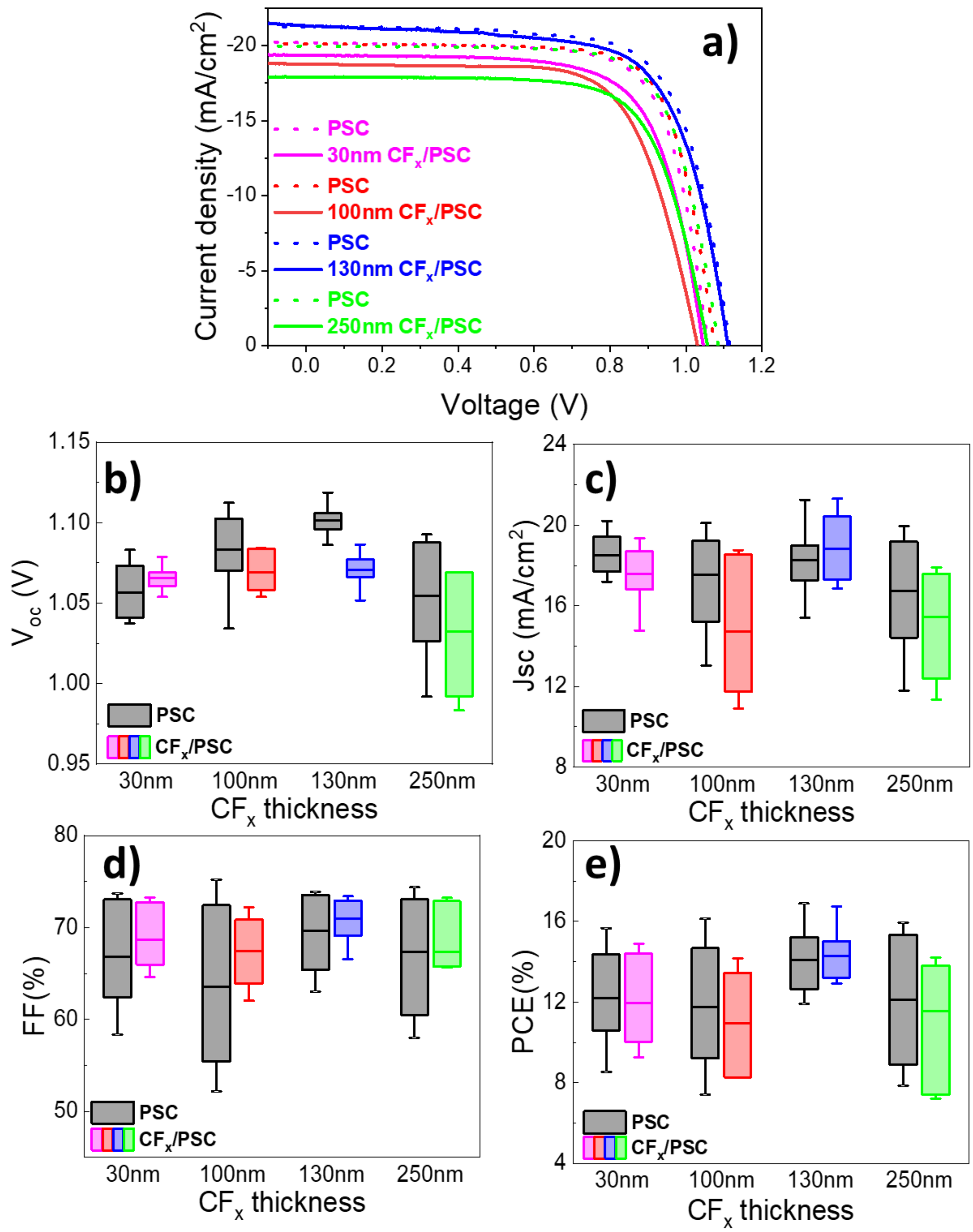

**Figure 2. Effect of the $CF_x$ thickness coating on the photovoltaic performance of the PSC.** a) Champion current density-voltage (J-V) curves under 1 sun (AM1.5G) illumination for $CF_x$-encapsulated perovskite solar cells with different $CF_x$ layer thicknesses (solid lines) compared to the uncoated reference (dotted line). b-e) Statistical analysis of photovoltaic parameters obtained from a batch of 40 different PSCs (approximately 10 for each configuration) using an ANOVA test [77]. The same batch used for the reference was later coated with $CF_x$. Parameters analyzed include: b) Open-circuit voltage ($V_{oc}$), c) Short circuit current density ($J_{sc}$), d) Fill Factor (FF), and e) Power Conversion Efficiency (PCE).

**Table 1**. Statistical photovoltaic parameters of reference and $CF_x$ encapsulated PSCs for the different $CF_x$ thicknesses. The table also shows the percentage of reduction of the mean value in every PV parameter. Champion values are presented in brackets.

| | $V_{OC}$ (V) | | | $J_{SC}$ (mA/cm$^2$) | | | FF (%) | | | PCE (%) | | |
|---|---|---|---|---|---|---|---|---|---|---|---|---|
| **Thickness (nm)** | **Ref.** | **- CFx** | **% Red.** | **Ref.** | **- CFx** | **% Red.** | **Ref.** | **- CFx** | **% Red.** | **Ref.** | **- CFx** | **% Red.** |
| **30** | 1.06 ± 0.02 [1.08] | 1.05 ± 0.02 [1.09] | -0.9 [0.3] | 18.5 ± 1.1 [20.2] | 17.6 ± 1.5 [19.3] | -5.0 [-4.2] | 66.8 ± 5.8 [73.7] | 62.3 ± 6.8 [71.5] | -6.7 [-2.9] | 12.2 ± 2.5 [15.7] | 10.9 ± 2.4 [14.2] | -10.3 [-9.1] |
| **100** | 1.08 ± 0.02 [1.11] | 1.06 ± 0.04 [1.10] | -2.4 [-0.9] | 17.5 ± 2.2 [20.1] | 14.7 ± 3.5 [18.8] | -16.1 [-6.7] | 63.5 ± 8.8 [75.2] | 59.9 ± 7.9 [69.5] | -5.8 [-7.6] | 11.7 ± 3.2 [16.1] | 9.8 ± 3.0 [13.4] | -16.5 [-16.7] |
| **130** | 1.10 ± 0.01 [1.12] | 1.06 ± 0.03 [1.11] | -3.6 [-0.9] | 18.2 ± 1.9 [21.2] | 18.8 ± 1.7 [21.3] | -3.2 [0.2] | 69.6 ± 4.1 [73.9] | 66.9 ± 4.9 [71.8] | -4.0 [-2.8] | 14.1 ± 1.6 [16.9] | 13.6 ± 1.5 [16.4] | -3.7 [-3.1] |
| **250** | 1.05 ± 0.04 [1.09] | 0.95 ± 0.12 [1.06] | -10.1 [-3.3] | 16.7 ± 3.0 [19.9] | 15.4 ± 2.8 [17.9] | -7.6 [-10.2] | 67.3 ± 6.8 [74.4] | 59.7 ± 16.7 [71.5] | -11.3 [-3.9] | 12.1 ± 3.5 [15.9] | 10.5 ± 3.7 [13.5] | -13.3 [-15.4] |

However, the best encapsulant thickness is obtained for an intermediate thickness of 130 nm. **Table 1** shows the PV statistical analysis, with the mean and the standard deviation, together with a reduction percentage (negative/positive for reduction/increase), where it can be noted that the encapsulation with the lowest reduction (below 4% for all the PV parameters) also corresponds to the 130 nm $CF_x$ encapsulated PSC. In addition, the champion parameters of the 130 nm $CF_x$ encapsulated device yield a PCE value of 16.4 % (compared to 16.9% non-encapsulated reference cell, supposing a 3 % of reduction) with $V_{oc}$ and $J_{sc}$ values changing from 1.11 to 1.10 V (1 %) and 21.2 to 21.3 mA/cm$^2$ (no reduction), respectively. Therefore, we will focus on this thickness to continue analyzing the stability of the cells.

## 2.3. Stability of the $CF_x$-encapsulated PSCs under simulated environments and compatibility with commercial encapsulants

**Figure 3** a)-d) display the evolution of the normalized PCE under continuous illumination (1 sun, AM1.5G) for the reference unencapsulated (black curve) and 130 nm $CF_x$-encapsulated (blue curve) PSCs under different environmental conditions, obtained by averaging the PCE out of 4 subcells of the same device. The complete set of photovoltaic parameters is shown in **Figure S4 and Table S1**. All stability tests were conducted in a custom-built environmental chamber, outside of the glovebox, allowing precise control over temperature, humidity, and gas atmosphere (see Experimental Section). This setup follows the recommendations outlined in the consensus statement on stability testing for perovskite solar cells. [79]

Figure 3 a) shows the evolution of the normalized PCE under continuous illumination in a simulated dry and inert atmosphere ($N_2$ flow). Both the reference and the $CF_x$-encapsulated devices exhibit a similar initial burn-in behavior during the first hour, marked by a rapid efficiency drop. After this period, the reference unencapsulated device undergoes continuous degradation, reaching 80% of its initial PCE ($T_{80}$) in approximately 8 hours. In contrast, the encapsulated device stabilizes after the burn-in phase and maintains nearly constant performance for more than 65 hours (see Table S1). Moreover, considering this burn-in period, the corrected $T_{S80}$ (80% decay of PCE excluding the rapid initial degradation called "burn-in" before the stabilized trend[79]) increases to >27 and >200 h (extrapolated) for reference and encapsulated devices, respectively. The stability slightly decreases for a 100% Relative Humidity (RH) nitrogen environment (Figure 3 b)), with the PCE for $CF_x$ encapsulated always higher than for the as-prepared cell. Photographs of the reference and $CF_x$ encapsulated devices in Figure 3 demonstrate the apparent degradation (yellowish coloration) at the edges of the gold layer for reference devices, especially clear when the samples are observed through the glass-side (highlighted with a yellow circle). In contrast, the appearance of encapsulated devices is practically unaltered, in agreement with the higher stability observed. The effect of a saturated, humid air atmosphere (generated by passing a flow of air through a water bubbler; see Experimental Section) is shown in Figure 3 c). Unlike the conditions in Figure 3 b), here the atmosphere contains both oxygen and moisture, which are known to synergistically accelerate perovskite degradation.[80,81] Accordingly, a fast and constant degradation is observed, with the PCE reaching 80% of its initial value ($T_{80}$) after only 0.5 and 2.5 h for the non-encapsulated and encapsulated cells, respectively (see Table S1). This direct comparison between panels b) and c) illustrates the detrimental effect of oxygen when combined with water vapor, leading to much faster degradation in air than in nitrogen, even at the same 100% RH.

$CF_x$-encapsulated PSCs also exhibited remarkable stability in $V_{oc}$ and FF parameters, especially under dry experimental conditions (Figure S4). Thus, the $CF_x$ encapsulated cells presented $T_{80}$ values around 65 and 32 h for dry and humid nitrogen environments, respectively (Table S1). Moreover, the corresponding $T_{S80}$ parameter with the fluorinated encapsulant was further improved by 7 and 2 times compared to the corresponding unencapsulated PSC one, reaching values of at least 200 h (dry) and 60 h (at 100% RH) (see Table S1). The test was then maintained for more than 10 days to observe the effect of prolonged exposure to moisture-saturated environments (**Figure S5**). Besides the initial faster degradation occurring for reference devices, the slope in the constant degradation region (t>180 h) is also higher for the reference (-0.14 $\pm$ 0.02 %/h) than for $CF_x$ encapsulated ones (-0.11 $\pm$ 0.02 %/h). Eventually, the PCEs of the encapsulated and reference devices decrease to ca. 50 and 35 % of their initial values after 270 h (Figure S5).

The last stability experiments were carried out at 50 °C (Figure 3 d)) for 2.5 days to simulate a realistic scenario of the rainy seasons and high temperatures reached during summer. In this case, the environment was nitrogen with 100% RH, similar to panel b) but at an elevated temperature. $CF_x$-encapsulated PSCs revealed an initial decrease (burn-in) during the first two hours.[82] The slightly higher burn-in losses observed in $CF_x$-encapsulated perovskite devices might be attributed to residual mechanical stress introduced during encapsulation. While $CF_x$ layers enhance the barrier properties and environmental stability of perovskite solar cells, their deposition process could induce stress at the interfaces. This hypothesis aligns with recent findings by other authors who have linked burn-in degradation to thermally induced stress, which can cause mechanical deformation within the device architecture.[82] Therefore, the residual stress associated with $CF_x$ encapsulation likely acts similarly to thermal stress, initiating early-stage degradation and contributing to a larger initial decay of the photovoltaic response. After the burn-in decay, almost constant reduction at a rate of 0.56 $\pm$ 0.01 %/h of the normalized PCE value until reaching around 60% after more than 50 h. The unencapsulated PSCs showed a more severe degradation, falling below 50% at the end of the test, with a much more marked trend slope during the first 20 hours. While $T_{80}$ times are similar for encapsulated and reference devices, 10 and 8 h, respectively, the $T_{S80}$ is much higher for the encapsulated device of approximately 40 h (note that, as the reference does not depict a clear burn-in region, $T_{S80}$ for unencapsulated devices remains as 12 h). The corresponding photographs show the evident degradation of the reference cells (see yellow circle in opposition to the encapsulated one). Importantly, the higher stability in panel d) compared to panel c) is attributed to the oxygen-free nitrogen atmosphere. This confirms that, even under high humidity and elevated temperatures, eliminating oxygen significantly slows degradation, highlighting the critical role of $O_2$ in perovskite instability. These trends are similar to those presented by thinner $CF_4$ plasma-assisted coating in a dry environment[74] and comparable to other hydrophobic fluoropolymer encapsulants, but with thicknesses of several µm.[25]

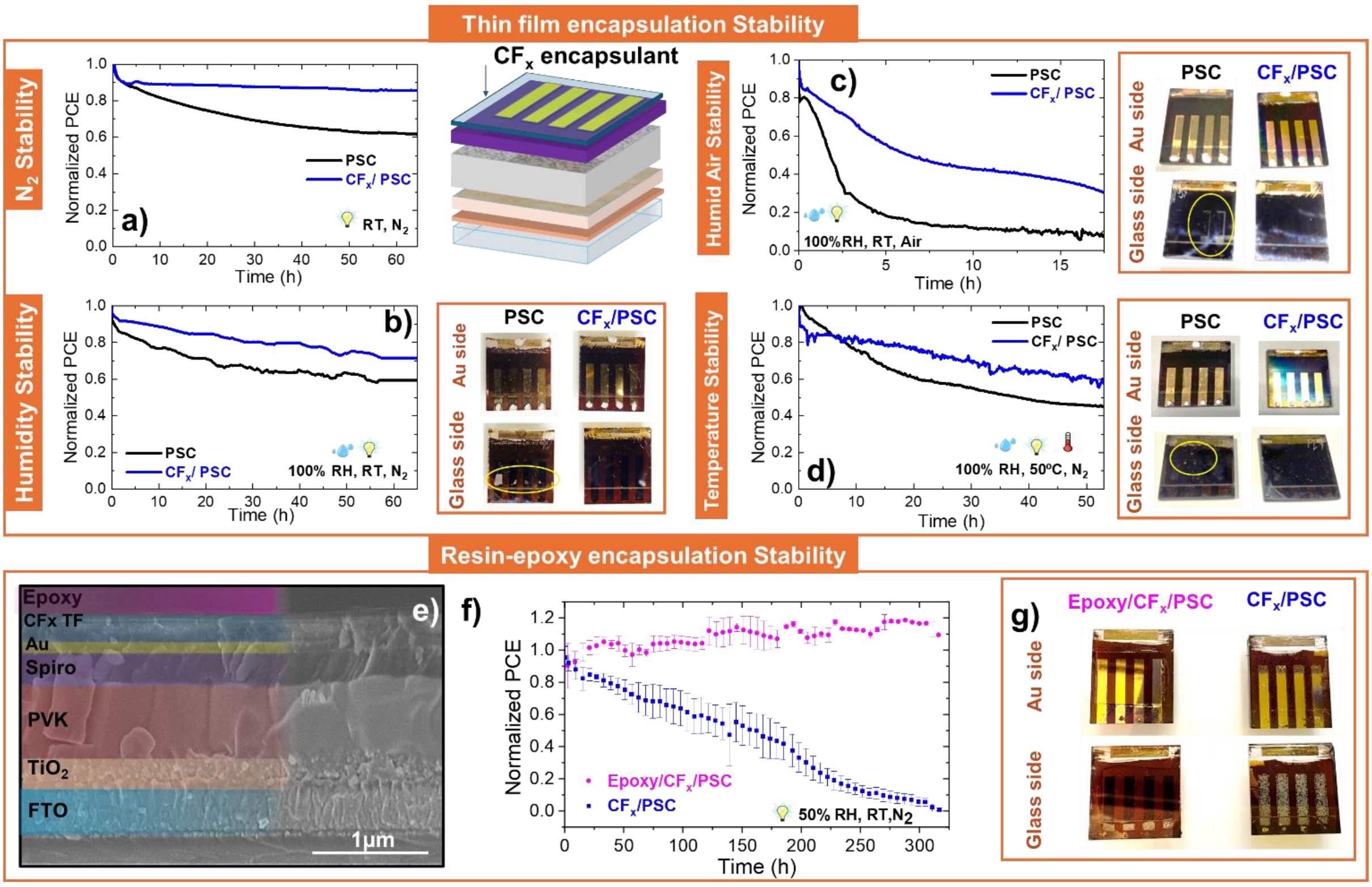

**Figure 3. Stability of the encapsulated PSCs ($CF_x$ (130 nm) and resin-epoxy/$CF_x$ (130 nm)) under different humid and temperature conditions**. **Top:** Normalized power conversion efficiency (PCE) evolution measured under constant illumination (1 sun – AM1.5G) in reverse scan using a scan rate of 0.02 V/s. The measurements were carried out at: a) dry ($N_2$ flow) atmosphere and 20ºC, b) 100% RH ($N_2$ flow through a water bubbler) and 22 °C, c) air flow through a water bubbler (to ensure the 100% RH); d) same conditions as in b) but at 50 °C. **Right side of the tests:** Scheme of the layer-by-layer structure of the PSC employing $CF_x$ thin film (a) and photographs of the reference and $CF_x$ encapsulated PSCs taken from Au and glass sides after the end of every test condition, where specific degradation points and brighter areas (associated with $PbI_2$ appearance) are highlighted with yellow circles (b-d). **Bottom:** e) Cross-section SEM micrograph and scheme of the layer-by-layer structure of the PSC employing $CF_x$ thin film as an interlayer between the PSC and the epoxy-commercial encapsulant. f) Normalized PCE evolution for an extended durability test carried out during 300 hours under 50% RH, $N_2$, and 20ºC. Refer to **Figure S6** in the Supporting Information section to follow the evolution of the absolute value of the PV parameters as well as cross-sectional images and WCA characterization before/after testing. g) photographs of the epoxy/$CF_x$ encapsulated PSC and $CF_x$ encapsulated PSCs taken from Au and glass sides after the end of the test.

While $CF_x$-coated devices exhibit reduced degradation, this improvement alone is insufficient to ensure long-term operation of hybrid PSC/D-TENG systems. To address this, we combined the hydrophobic $CF_x$ layer with a commercial UV-curable epoxy resin (Ossila E132), as shown in Figure 3 (e–g). Although epoxy resins like E132 are commonly used as encapsulants for perovskite solar cells due to their excellent barrier properties against moisture and oxygen,[83] their application is highly dependent on the specific device architecture. In particular, some UV-curable epoxies can chemically interact with Spiro-OMeTAD, one of the most widely employed hole transport layer in perovskite technology. This incompatibility arises from chemical interactions between the methoxy-containing compounds in Spiro-OMeTAD, leading to rapid degradation of the HTL and full suppression of photovoltaic response.[23,84,85] We experimentally confirmed this effect in control devices encapsulated directly with E132, which showed no photovoltaic activity. Hence, a key outcome of our work is that the introduction of a $CF_x$ interlayer enables the use of epoxy encapsulants in devices employing Spiro-OMeTAD by preventing direct chemical interaction between the resin and the active layers. It is worth stressing here that our approach does not aim to replace conventional bulk encapsulants such as EVA or POE; rather, $CF_x$ acts as an interlayer that prevents chemical interactions between liquid-phase encapsulants and sensitive layers such as Spiro-OMeTAD. This strategy is especially relevant for n-i-p perovskite cells, where epoxy resins typically induce severe degradation.

Hence, a key outcome of our work is that the introduction of a $CF_x$ interlayer enables the use of epoxy encapsulants in devices employing Spiro-OMeTAD, by preventing direct chemical interaction between the resin and the active layers. The $CF_x$ layer acts as a chemically inert and conformal barrier, allowing successful encapsulation without structural or functional degradation of the perovskite or HTL, as evidenced by the cross-sectional SEM image in Figure 3 e). Remarkably, the combination of $CF_x$ and epoxy encapsulation leads to significantly enhanced device stability, as it can be seen in the stability curves of Figure 3 f), which represent average values obtained from 4 subcells per device and their standard deviations. Epoxy/$CF_x$-encapsulated PSCs maintain over 90% of their initial efficiency under RT humid nitrogen flow (50 % RH) for more than 300 hours (Figure 3 f), outperforming the conditions and duration tested in Figure 3 b). By contrast, $CF_x$/PSC devices degrade rapidly, falling below 80% within 20 hours and reaching zero after 300 hours. Although the epoxy/$CF_x$ device shows an initial sharp decline, it later recovers and stabilizes above 90% after 300 hours. In epoxy/$CF_x$ encapsulated samples, certain fluctuations in the performance evolution can be observed. These are attributed to variability in individual subcells, which may arise from imperfect $CF_x$ coverage or small particles at the interface introduced during processing in non-cleanroom environments. Such imperfections can lead to localized infiltration of the liquid epoxy resin, causing slight electrical instabilities in some electrodes. Although this results in less smooth curves, it provides a realistic representation of the device behavior under practical fabrication conditions. The photographs in Figure 3 g) show a higher degradation for the reference and the $CF_x$-encapsulated cell (significantly degraded in the brighter areas) than for the epoxy/$CF_x$ encapsulation with an unaltered brownish color. The evolution of other key photovoltaic parameters is presented in **Figure S6.** It is noteworthy that the normalized FF and $V_{oc}$ parameters remain stable until 200 h for both epoxy/$CF_x$/PSC and $CF_x$/PSC, with similar $V_{oc}$ values and slightly higher FF ones for the latter case. Cross-sectional SEM images and wetting evolution pictures of epoxy/$CF_x$/PSC and $CF_x$/PSC before and after the aging test (Figure S6) confirm that the introduction of a hydrophobic $CF_x$ interlayer enables the use of epoxy encapsulants in devices employing Spiro-OMeTAD by preventing direct chemical interaction between the resin and the active layers. Particularly, SEM micrographs of the aged samples show that the $CF_x$ encapsulation remains continuous and defect-free, with no cracks, pinholes, or delamination, even near mechanically and chemically critical interfaces such as $CF_x$/Au electrode and CFx/Spiro-OMeTAD HTL.

To further assess the robustness of the encapsulation strategy under harsher environmental conditions, **Figure 4** a) presents the stability evolution of PSCs exposed to air at 80% relative humidity, first at room temperature, followed by a thermal stress stage at 100 °C. Under these conditions, both the unencapsulated reference and the $CF_x$-only encapsulated devices degraded completely within 60 hours, confirming their poor tolerance to moisture and oxygen. In contrast, the epoxy/$CF_x$ encapsulated PSC exhibited a markedly improved durability, retaining approximately 40% of its initial efficiency after 60 hours at 80% RH and RT. Upon increasing the temperature to 100 °C (while maintaining the same humid air environment), the device continued to decline until reaching a plateau around 25% after ~80 hours, where it remained approximately stable until the end of the test at ~170 hours. This behavior highlights the synergistic effect of the $CF_x$ interlayer and epoxy encapsulation in mitigating degradation pathways even under combined moisture and thermal stress, although the performance loss is more pronounced than under inert conditions (c.f. Figure 3 f)).

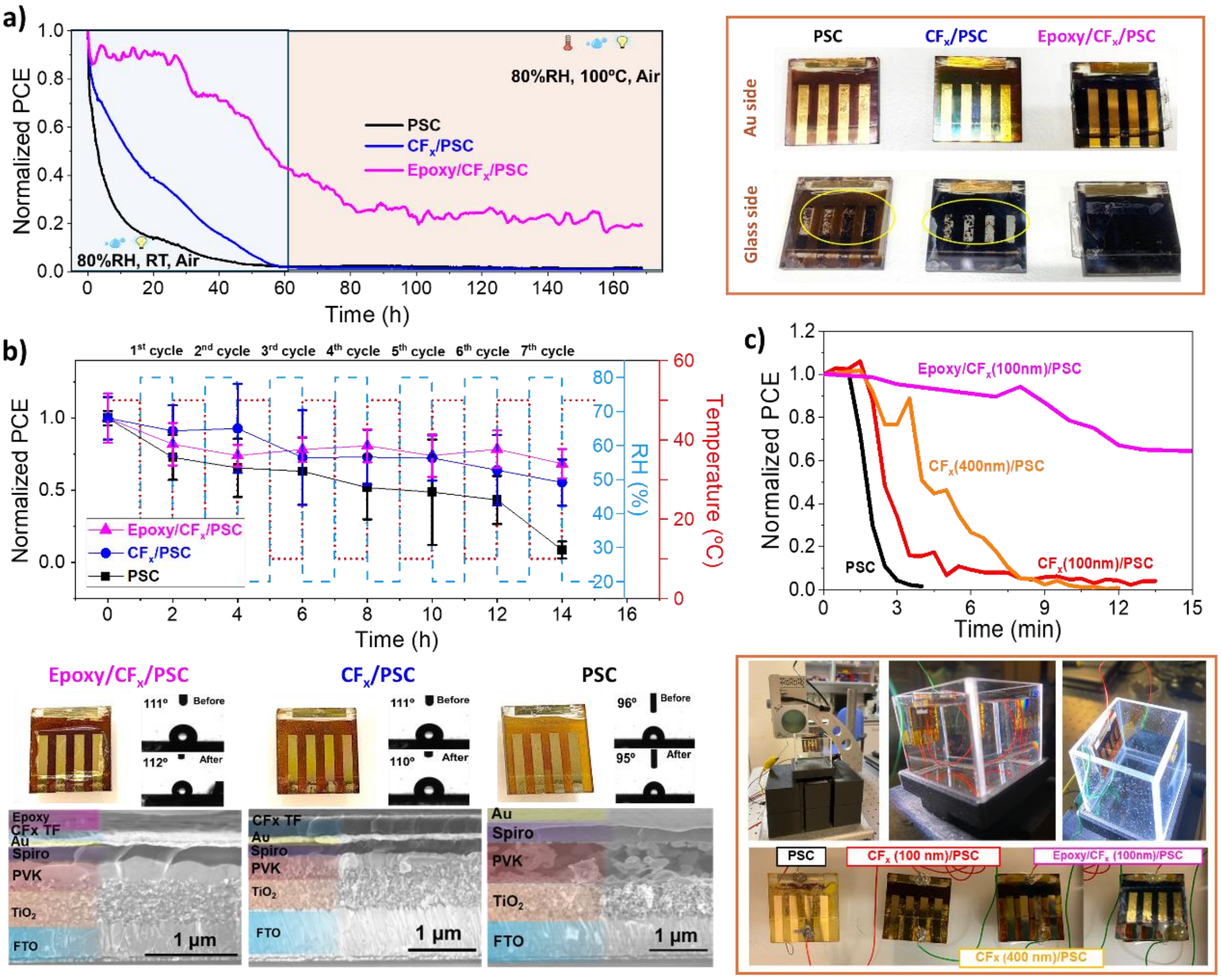

**Figure 4. Stability of the encapsulated PSCs ($CF_x$ and epoxy/$CF_x$) under illumination at high humidity, cycling humidity and temperature, and liquid water immersion.** a) **Left**: Evolution of normalized PCE measured under constant illumination (1 sun, AM1.5G) in reverse scan at a rate of 0.02 $V \cdot s^{-1}$. Tests were performed at 80% RH in continuous airflow (RH controlled through a water bubbler) at 20 °C for 60 h, followed by 100 °C for 100 h. **Right**: Images of the reference, $CF_x$-encapsulated, and epoxy/$CF_x$-encapsulated PSCs (taken from Au and glass sides) after the stability test, with degradation points and visible degradation in active areas highlighted in yellow. b) **Top:** Normalized PCE for reference, 130 nm thick $CF_x$ encapsulated PSC with, and epoxy/130nm $CF_x$ under repeated stress cycles (1 h at 50ºC and 20% RH + 1 h at 10ºC and 80% RH). **Bottom**: Images of the same devices than top panel, along with wetting and cross-sectional SEM images before and after degradation testing. c) Immersion test in water. **Top**: Normalized PCE of reference PSC, $CF_x$-encapsulated PSCs (100 nm and 400 nm), and epoxy/$CF_x$-encapsulated PSC for 15 min immersion. **Bottom**: Setup for photovoltaic characterization under immersion (solar simulator rotated 90° and water cuvette with PSCs attached to one wall) and photographs of PSCs after the test.

The extended stability under such aggressive conditions demonstrates the potential of this encapsulation approach for real-world applications where devices may face simultaneous humidity and temperature challenges. To complement these static damp-heat tests and better emulate real-world stressors, we implemented a thermal-humidity cycling protocol, as recommended in PV reliability standards and recent perovskite stability studies. Such dynamic conditions are known to activate intrinsic degradation pathways, primarily ion migration and halide phase segregation, rather than macroscopic failure, even in well-encapsulated devices.[86] This accelerated protocol, designed to emulate harsh outdoor fluctuations in air, consisted of alternating 1 h stages at 80% relative humidity and 10 °C, followed by 1 h at 20% relative humidity and 50 °C, repeated continuously for up to 14 h of cycling. Under these conditions, the non-encapsulated PSC failed rapidly, reaching zero efficiency within the first 14 h. In contrast, both encapsulated devices showed substantially higher resilience, retaining more than 50% and 70% of their initial PCE, for $CF_x$ and Epoxy/$CF_x$ encapsulated PSC, over the same period. Their evolution was similar, showing a gradual decline with the number of cycles rather than abrupt failure. These results underline the fact that, while the epoxy resin significantly enhances stability under continuous illumination, high humidity, and elevated temperature, and effectively limits oxygen ingress, thermal-humidity stress cycling activates different degradation pathways in which the epoxy layer plays a more limited role. As reported in the literature, such failures are typically driven by nanoscale processes such as ion migration and phase segregation, exacerbated by repeated thermal excursions, rather than by macroscopic cracking or delamination.[86,87] SEM analysis confirmed that no cracks or delamination occurred, even near critical interfaces such as $CF_x$/Au and $CF_x$/Spiro-OMeTAD whereas the stability of the water contact angles values measured before and after the durability tests (Figure 4 b bottom or Figure S6 b), supporting the hypothesis that degradation is driven by nanoscale processes rather than structural failure. The superior stability observed for epoxy/CFx-encapsulated devices can be partially attributed to the remarkable mechanical properties of the $CF_x$ interlayer, as detailed in Figure S3. Nanoindentation measurements revealed a Young's modulus of 3.7 $\pm$ 0.2 GPa and a hardness of 0.23 $\pm$ 0.01 GPa. This enhanced rigidity, combined with the inherent polymeric flexibility of the CFx coating, suggests an improved resistance to plastic deformation. Such behavior is consistent with the formation of highly crosslinked polymeric networks induced by plasma polymerization during fabrication, which confer structural integrity while allowing limited plastic accommodation under stress. Consequently, the $CF_x$ layer acts not only as a chemically inert barrier but also as a mechanically robust interface capable of mitigating stress concentrations and preserving encapsulation integrity over extended aging periods (see also Figure S6).

Beyond enabling the combination with commercial epoxy encapsulants (for instance, to avoid interaction with Spiro-OMeTAD layer), the $CF_x$ layer itself exhibits promising barrier properties. To explore its intrinsic encapsulation capability under extreme conditions, we assessed the performance of PSCs encapsulated solely with $CF_x$ when exposed to liquid water. This approach isolates the contribution of $CF_x$ as a standalone protective layer. **Figure S7** presents the J-V curves comparing the reference and $CF_x$/PSCs measured after being submerged in water after 180 s, demonstrating the superior stability of the encapsulated cells, (including experiments for the lowest thickness to ensure complete coverage, i.e., 100 nm and for the thickest transparent layers, i.e., 400 nm of $CF_x$). This preliminary experiment demonstrated that the encapsulated samples withstand immersion in liquid water. To further investigate, we designed an experiment to monitor the J-V curves under simultaneous illumination and immersion. The experimental setup is illustrated in Figure 4 c-bottom). Figure 4 c-top) highlights the enhanced water resistance by depicting the evolution of the normalized PCE under these simultaneous illumination and immersion conditions. Thus, after the first minute, degradation mechanisms are triggered abruptly, and after 2-3 min the unencapsulated PSC completely deteriorates ($T_{80}$ = 80 s). However, with the thinner $CF_x$ encapsulant (100 nm, red curve), the beginning of the degradation curve takes twice as long as reference devices and presents a higher $T_{80}$ of more than 120 s. Moreover, higher waterproof protection was obtained for the 400 nm $CF_x$ thickness as the normalized PCE showed an improved $T_{80}$ of ca. 200 s. It is important to note that the changes in the optical paths as light absorptions/reflections in the water and cuvette were not considered for estimating the photovoltaic parameters, which can affect the actual PCE values (especially important for the non-normalized J-V curves in Figure S7). Nevertheless, the most significant result is the demonstrated stability in water immersion and under the epoxy/$CF_x$ encapsulated PSC illumination for more than 15 min, providing a $T_{80}$ value of 600 s. The pictures in Figure 4 c) show the brownish appearance of the encapsulated cells compared to the total turn to yellow for the as-prepared one.

These outstanding results are comparable to recent works where water immersion has been ensured for 24 hours with several hundred microns-thick encapsulants without simultaneous solar illumination.[28,88,89] It is worth noting that the $CF_x$ encapsulation of the perovskite solar cells was performed in a single step on the gold electrode side. Therefore, further improvements in durability are likely achievable by repeating the deposition process or by making slight adjustments to the geometrical conditions to enable sealing from the top, bottom, and edges. The same reasoning applies to the combination with standard encapsulants (such as Ossila), as effective edge sealing has been shown to be a critical factor in resisting moisture or water exposure, a consideration beyond the scope of this study. Also, the same encapsulation protocol can be extended to alternative halide perovskite compositions or crystalline structures to enhance the overall durability of the cells.

## 2.4. Thin film-based drop triboelectric nanogenerator and hybridization with halide perovskite solar cell.

The outstanding stability of the epoxy/$CF_x$ encapsulated PSC opens the path toward its hybridization with drop triboelectric nanogenerators (D-TENGs). The first step is demonstrating the $CF_x$ triboelectric performance and solid-liquid nanogenerators. Fluorinated polymers are one of the materials of choice for their implementation in both solid-solid and solid-liquid TENGs due to their appealing electron affinity, capability for long surface charge durability, high dielectric constants, and tunable hydrophobicity by well-established patterning routes.[50–53,90] However, the topic is dominated by bulk polymers and foils, with a reduced presence of thin polymer films.[4,52,53] Figure 5 summarizes the performance of a $CF_x$ triboelectric layer for solid-liquid configurations to different droplet dripping, i.e., Milli-Q water and rainwater collected in Seville (February 2024). Please note that, for this example, we have further optimized the chemical composition of the plasma polymer while keeping the thickness at ca. 130 nm. Thus, although the fluorine content in this $CF_x$ layer is lower than that used as an encapsulant (see Figure S2 for XPS analysis comparison), the optimized chemical functionalization, rich in highly electronegative species such as -$CF_2$ and -$CF_3$ (which together account for 36.4% of the atomic surface content), produces an enhanced triboelectric output. The shape of the $V_{out}$ curve (Figure 5a) depicts a characteristic profile for rainwater, i.e., a sharp positive peak instantaneously produced during the spreading of the droplet and first contact with the top electrode ($t_0$) followed by a shallower and broader negative peak corresponding to the recoiling of the drop, ending when the drop loses contact with the electrode ($t_f$ in Figure 5 a)).[53,71] It is worth highlighting that the peak of Figure 5 a) of 110 V stands in sharp contrast to the remaining values of Figure 5, which are approximately around 60 V. This peak has been selected as the champion peak, obtained after multiple droplet impacts on the surface due to cumulative triboelectric charge transfer during each droplet event (see **Figures S8** to **S13**).

The intensity and duration of the peaks depend on the type of water. For the fixed resistance load of ca. 10 MΩ, higher positive peaks are obtained for rainwater than for Milli-Q water, reaching charge-transferred values for a single drop of 0.25 nC (Figures S8 and S9). These results are aligned with previously reported for D-TENGs.[50,51,53] Best performance includes open-circuit voltage peaks up to 110 V. The maximum power output was estimated at ca. 4 mW/cm$^2$ for a device with an Au top electrode (0.1 x 0.5 cm$^2$), rainwater, and 500 kΩ of load resistance ($R_{load}$) as presented in Figure 5b). Additional parameters, including mean power, energy per cycle, peak voltages ($V^+/V^-$), and peak shape ($FWHM^+/FWHM^-$), are provided in Figure S8. Contrary to the estimation of the PCE in the solar cells, there is no standard for evaluating the efficiency of D-TENGs (see the instantaneous peak power and energy and the equations used for their calculation in Table S2). However, the values in Figure 5 a-b) are competitive compared to the few alternative thin film TENGs, such as those reported by Q. Liang *et al*.[91] for a thin film of PTFE (ca. 300 nm in thickness) and multi-bottom electrode architecture, Chung *et al*.[92] for a fluorinated liquid-infused surface for two bottom electrodes, and by Lee *et al*.[93] for a hierarchical plasma fluorocarbon thin film of 100 nm.

The reproducibility of the obtained peaks depends strongly on the relative impact position of the falling drops with respect to the electrodes and the hydrophobicity of the triboelectric surface, the latter controlling the presence of remaining water on the surface for consecutive droplets. The variation is also a consequence of the charging of the triboelectric surface by contact with the first drops.[71] Figure 5 c) illustrates the sensitivity to droplet impact position, showing that the output voltage can vary by approximately 70% between direct electrode impact and sliding impact on the $CF_x$ surface. Furthermore, Figures 5 d – f) and Figure S10 present the influence of additional parameters governing triboelectric dynamics, with optimal responses exceeding 50 V under high-frequency conditions (3 Hz), a 45° tilt, and droplet volumes above 35 µL. A practical configuration was also identified, involving a 30 cm drop height and electrically ungrounded droplet dispensers.

A practical strategy to mitigate the positional dependence is to reduce electrode size and arrange multiple electrodes across the triboelectric surface, thereby increasing the probability of direct droplet-electrode interactions. To demonstrate this concept, we evaluated a $CF_x$-based D-TENG using a submillimetric electrode array, following a previously reported configuration.[71] **Figure S11** shows the output voltage of the $CF_x$ D-TENG in a 2×2 array configuration. The response was highly non-uniform: one electrode reached peak voltages of approximately 20 V, another barely achieved 2 V, and the remaining two were below 0.5 V. The two electrodes with the highest output correspond to the upper positions when the sample was tilted at 45°, consistent with previous observations.[71] This behavior is likely influenced by droplet impact position and water accumulation, as supported by Figure 5 c) and **Video S1**), which demonstrates the strong dependence of signal amplitude on impact location. Additional factors such as poor adhesion of the narrow gold electrodes, uneven wetting, or variations in contact resistance may also contribute to the pronounced differences observed, indicating the need for further optimization to achieve uniform performance across the array. Despite these issues with the submillimetric arrays, the standard top-bottom electrode configuration demonstrated a high resilience. Figure 5 g) presents a voltage signal retention of over 85% of the initial output after more than 17,000 drop events, approaching the values reported in the state of the art summarized in **Table S3** of the Supporting Information. In addition, durability tests in solid-solid mode against an ITO/PET counter electrode in Figure 5 h) confirmed the long-term robustness of the $CF_x$ coating, retaining the 70% of the initial output after more than 100,000 impacts (**Video S2**). The response under water-type cycling (rainwater and Milli-Q water), extended drop-impact tests, and surface charging phenomena are shown in Figures S12 and S13.

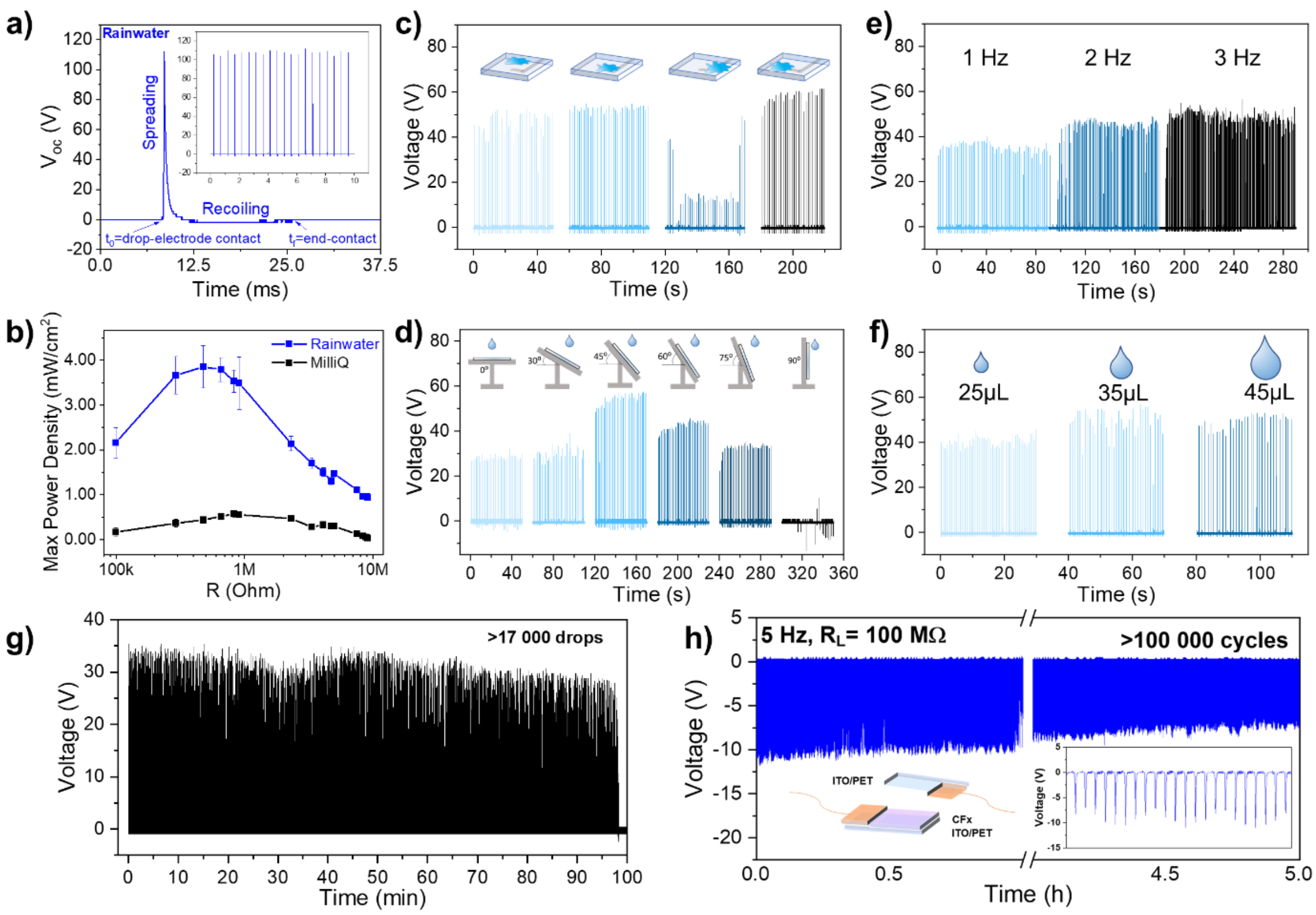

**Figure 5. Triboelectric characterization of $CF_x$ surfaces in D-TENG devices**. a) Open-circuit voltage signal of an optimized single raindrop, with detailed droplet dynamics and an inset showing the output signal for multiple droplets. b) Maximum power density (the square of the voltage divided by the load resistance) as a function of load resistance for dripping tests using both rainwater and Milli-Q water. c-f) Open-circuit voltage output under rainwater dripping for different experimental conditions: c) droplet impact position, d) platform tilt angle, e) droplet impact frequency, and f) droplet volume. Note that the rest of the parameters have remained constant (45º tilt, 35 µL, 30 cm height, ungrounded tip, and 1 Hz of frequency). g) Durability test for continuous dripping of Milli-Q water for more than 17,000 drops. h) Solid-solid output voltage signal during a mechanical test with more than 100,000 impacts against an ITO/PET counter electrode at 100 MΩ load resistance and 5 Hz tapping frequency (inset shows individual voltage signals).

The selected architecture enables the dual functionality of the $CF_x$ layer, acting both as an encapsulant for the metal halide perovskite and as a triboelectric surface for the D-TENG. This design is conceptually similar to the configuration proposed by Wang *et al*. for instantaneous droplet energy harvesting,[53] where the triboelectric layer is embedded between an extended bottom electrode (FTO) and a thin top metallic electrode (Au). However, in our case, the $CF_x$ layer remains supported on a glass substrate, as shown in the final architecture of the hybrid cell (Figure 6 a)). Although the present architecture requires separate $CF_x$ layers for encapsulation and triboelectric functions (one protecting the back contact and the other deposited on the cell front glass), this strategy addresses the asymmetric vulnerability of the device, where the active multilayer stack is significantly more sensitive to environmental degradation than the substrate side. Nevertheless, this approach remains compatible with future top-illumination architectures employing transparent electrodes, in which a single $CF_x$ layer could act simultaneously as both an encapsulant and a triboelectric interface. Such dual-function integration would streamline fabrication, reduce material usage, and enable the development of more sustainable hybrid devices.

The individual mechanisms for photovoltaic conversion (Figure 6 b)) and droplet-driven triboelectric generation (Figure 6 d)) are shown separately for clarity. In the photovoltaic mode, light absorption in the perovskite layer generates electron-hole pairs, which are subsequently separated and extracted through selective contacts (ETL and HTL), driving current through the external circuit. The mechanism for voltage generation by contact electrification and electrostatic induction in d) is based on previous models reported by the groups of Zuankai Wang (Xu *et al*. [56] and Xu *et al*.[[58]]) and previous studies from Zhong Lin Wang group in 2014 [55]. During D-TENG operation, when the drop contacts the CFx surface, an electric double layer (EDL) forms at the liquid-solid interface. Following the "Wang" model, the formation of the EDL takes two steps: first, the transfer of electrons from the drop to the surface (which presents, in this case, a high electron affinity), and secondly, ionic adsorption from the liquid onto the negatively charged surface, stabilizing the interfacial charge distribution. The formation of the EDL at the liquid-solid interfaces produces negative charges in the liquid that can be transferred to the top electrode when the drop contacts it. As the droplet spreads, moves, and detaches from the electrode, the dynamic modulation of contact area and capacitance alters the EDL configuration, which induces alternating voltage/current peaks across the external circuit by electrostatic induction. Finally, Figure 6 c) illustrates the hybrid mode, where simultaneous illumination and droplet impact enable concurrent harvesting of solar and triboelectric energy, managed through a shared electrode and an integrated energy management circuit inspired from previous models reported by Bao *et al*. [69].

**Table 2** summarizes the different device configurations and encapsulation strategies employed to demonstrate the hybrid concept. These are grouped into two sets: one tested for durability and another for proof-of-concept demonstrations, corresponding to Figures 6 and 7, respectively. A detailed description of cell preparation is provided in Section S14 of the Supporting Information.

**Table 2.** Photovoltaic parameters of hybrid PSC/D-TENGs evaluated for durability and multisource energy harvesting for the champion cell.

| Device | $V_{oc}$ (V) | $J_{sc}$ (mA/cm$^2$) | FF(%) | PCE (%) |
|---|---|---|---|---|
| **PSC/D-TENG PV parameters for the as-prepared device for durability tests** | | | | |
| **PSC** | 1.10 | 18.25 | 71.0 | 14.1 |
| **$CF_x$ (130nm)/PSC** | 1.06 | 18.82 | 66.9 | 13.6 |
| **epoxy/$CF_x$(130nm)/PSC/CFx (130nm) (Fig. 6-7)*** | 1.11 | 8.46 | 61.0 | 11.4 |
| **PSC/D-TENG PV parameters for the as-prepared device for multisource energy harvesting** | | | | |
| **PSC** | 1.10 | 21.23 | 74.0 | 17.1 |
| **$CF_x$ (130nm)/PSC** | 1.12 | 22.47 | 71.0 | 17.9 |
| **epoxy/$CF_x$(130nm)/PSC/CFx (130nm) (Fig. 6)*** | 1.09 | 11.56 | 62.7 | 15.7 |

*Hybrid cell while dripping and solar simulator at 0.5 suns

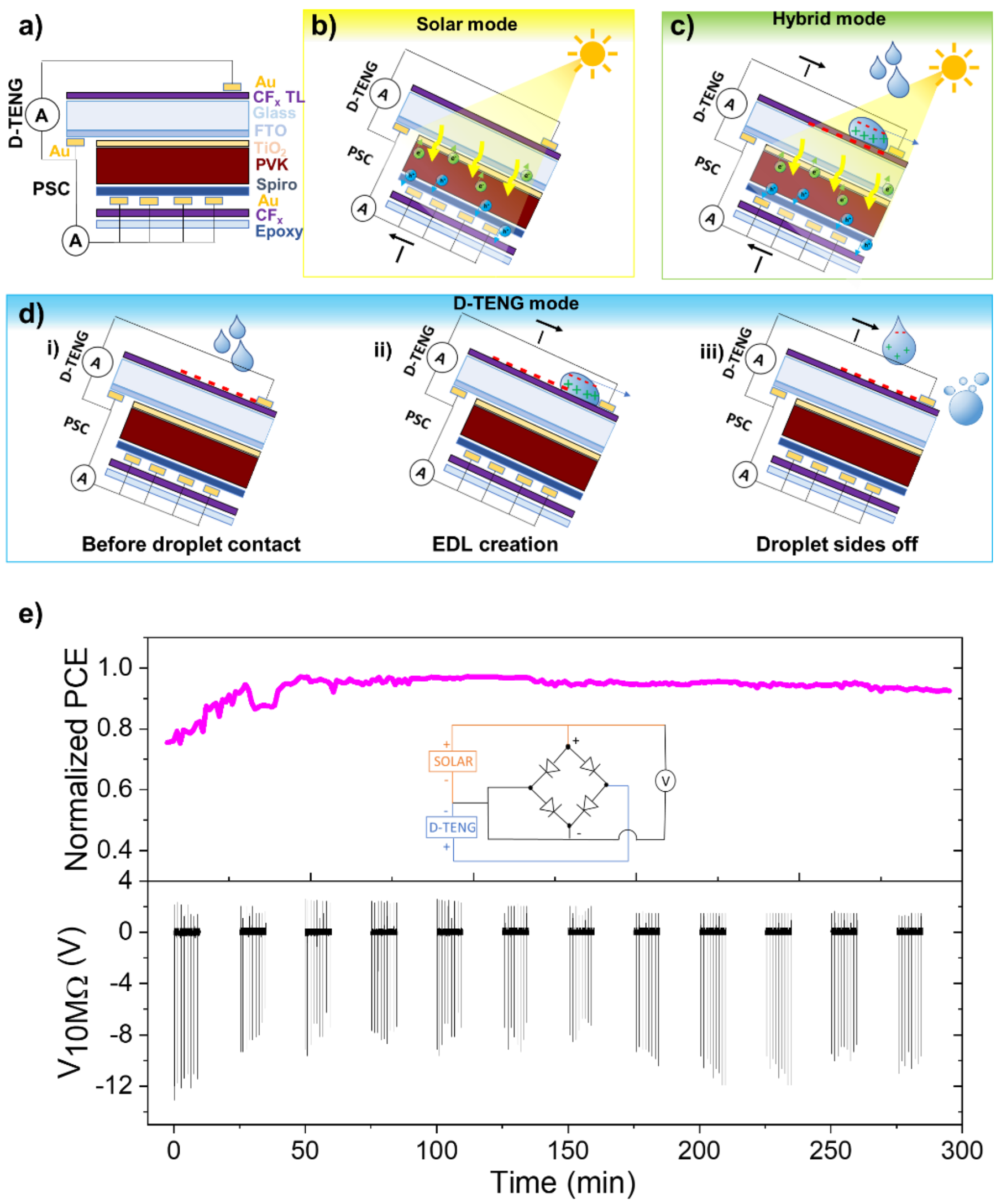

**Figure 6. Durable rain and solar energy harvesters enabled by $CF_x$ thin film encapsulants**. a) Scheme of the layer-by-layer assembly of the hybrid PSC / D-TENG device, including the shared FTO electrode for both harvesters. b)-d) Charge generation and extraction mechanisms for the individual harvesters for photovoltaic (b), droplet energy harvesting (d), and hybrid mode (c). The mechanisms for voltage generation by contact electrification and electrostatic induction in d) are based on previous models reported by Xu *et al.* [56], Xu *et al*. [58] and Lin *et al.*[55] e) Normalized PCE for 0.5 suns illumination and dripping raindrops on devices positioned at a 45° angle. The top panel shows PCE under periodic raindrop dripping at 1 Hz, while the bottom panel displays the corresponding voltage output for the D-TENG counterpart. The device includes equivalent 130 nm $CF_x$ layers as the encapsulant and triboelectric surfaces. The $CF_x$ was combined with epoxy resin for improved encapsulation. The experiment was carried out for rain droplets of 37 µL falling from a fixed height of 30 cm with the hybrid device forming 45°. The inset scheme shows the energy management circuit employed for the simultaneous characterization of both the solar cell and D-TENG.

The integration of the $CF_x$ D-TENG with the PSC follows the schematic shown in Figure 6 a) for a $CF_x$ thickness of 130 nm of the more crosslinked nature, i.e., equivalent to the encapsulant configuration (Figure 1 e)). The Au electrode was deposited on top of the $CF_x$ layer, while the FTO of the cell served as the bottom electrode of the D-TENG. The bottom Au electrode in contact with the HTL was protected using an epoxy/$CF_x$ combination. A power management circuit was employed for simultaneous characterization of the PSC and the D-TENG, as illustrated in the inset of Figure 6 e). For these experiments, 37 µL raindrops were released from a fixed height of 30 cm onto the hybrid device positioned at a 45° tilt. The solar simulator was also tilted at 45° to ensure normal light incidence on the surface (see **Video S3** and **Figure S14**). Notably, the $CF_x$ coating exhibits high transparency (Figure 1 a), inset); the 130 nm layer acts as an anti-reflective coating, enabling its integration on top of the solar cells while increasing light transmission. A detailed description of the light-intensity calibration procedure and the characterization of both the photovoltaic and combined photovoltaic–D-TENG systems is provided in Supporting Information **Section S15**.

The initial PCE of the fully encapsulated PSC measured under standard conditions is 11.4 % (note this is a different batch from the champion cells discussed in Figure 2), in comparison with the 14.1 % of the non-encapsulated cell (see Table 2). The observed discrepancy between the photovoltaic parameters is mainly related to the complexity of the workflow and our effort to minimize the number of $CF_x$ deposition batches (130 nm thick), which required selecting different cells for different experiments. One cell was chosen for the multisource energy harvesting test due to its superior performance, while the other was used for the durability test, where its initial PCE was considered acceptable. This context explains the variability observed in Table 2. The hybrid device was tested under continuous illumination (0.5 suns) while subjected to dripping cycles consisting of 10 minutes of raindrop impact (frequency: 1 Hz) followed by 15 minutes of rest. Although these conditions do not correspond to standard rain erosion tests, the dripping frequency applied here is two orders of magnitude higher than that expected under moderate rain and one order higher than heavy rain conditions.[94] It is important to note that previous studies in our laboratory have demonstrated the excellent rain erosion resistance of $CF_x$ surfaces. As reported in reference [95], aging tests were previously performed for these $CF_x$ coatings involving water droplets with an average diameter of 260 µm impacting at different velocities. For $CF_x$ coatings, more than 200 impacts at 165 m/s (much higher than the terminal velocity of raindrops, i.e., <10 m/s[94]) were required to produce visible damage, and over 2000 impacts were needed to induce a wettability change (15% variation in water contact angle).

The normalized PCE evolution and the voltage signal of the D-TENG upon periodically dripping are shown in Figure 6 e). Note that the real PCE value is not easily extrapolated during the simultaneous characterization for several reasons detailed in Supporting Information S15. In any case, the evolution of the normalized PCE can be used to estimate the stability of the devices under such aggressive conditions. As shown in Figure 6 e), the PCE continuously increases during the first 50 min of the experiment and after 2 cycles of 10 minutes of dripping (1200 drops). This enhancement of the PCE can be linked to changes in the refractive index during water dripping,[96] temperature equilibration, or other intrinsic stabilization mechanisms of the perovskite.[97] The hybrid system also profits from the enhanced durability achieved by the epoxy/$CF_x$ encapsulation. Thus, the PCE reaches an unaltered plateau upon dripping for over 4 hours, with the D-TENG counterpart reaching up to 12 V of peak voltage (Figure 6 e)-bottom) driven by the D-TENG. The variations in the triboelectric output voltage are related to the small fluctuations of the impacting drops on the top electrode (see also Figure 5 c). We have previously reported that even small displacements on the order of 0.5 mm can significantly alter the effect of the spreading dynamics on the triboelectric response, leading to voltage signal variations of up to 50%.[71] Although we employ here an automatic peristaltic pump to control the frequency and volume of the drops, the relative position between the thin top electrode and the syringe slightly varies from one experiment to another due to the 45º tilting of the holder, environmental vibrations, and airflows, and it is corrected manually. In the configuration for the simultaneous illumination and dripping, the initial PCE of the PSC is 11.45 % with a $J_{sc}$ of 8.46 mA/cm$^2$, considering the balance between the overestimation in terms of the active area and the underestimation of the light that reaches the PSC (for an active area of 0.36 cm$^2$ and 0.5 sun of illumination). These values yield above 90% of the measured PCE under standard conditions.

These results pave the way for developing resilient hybrid solid-liquid triboelectric nanogenerators and halide perovskite solar cells. However, significant efforts are still needed from both communities to build on these initial findings. Here we present a proof-of-concept, but achieving long-term stability will require substantial advancements at various levels. First, much more stable halide perovskite photo-absorbents need to be engineered. We hypothesize that implementing low-dimensional MAPI and exploring alternative compositions of the perovskite [9,13,15,36] and selective layers can help in this regard. In light of the latest findings in the literature,[71] we anticipate a slight decrease in the performance of the D-TENG after prolonged drop impacts, primarily due to charge saturation on the surface. The devices can be restored by drying and resetting the triboelectric surface, but enhancing the understanding of triboelectric charge generation and transport in solid-liquid interactions is essential to improve durability and performance. Additionally, the top-performing D-TENG architectures reported to date often expose the top electrodes to droplet impacts, which can quickly erode and corrode delicate thin metal materials. Therefore, advances in developing transparent, durable, and robust electrodes are critical.

Together to address the issues of durability and reproducibility, a significant challenge in the topic of multisource energy harvesting is to develop energy management circuits able to address the high voltages, low current, and short AC peak duration characteristic of triboelectric and piezoelectric nanogenerators, at the same time that DC low voltage and relatively higher current are generated by solar cells. Aiming to offer a proof-of-concept for actual multisource energy harvesting, we have evaluated different configurations for self-charging devices, including a hybrid PSC/D-TENG combined with a supercapacitor (see the PV parameters of this system in Table 2, the triboelectric output is equivalent to the system in Figure 6). This hybrid device has been tested sequentially under MilliQ and 0.5 suns and simultaneously employing alternative energy management circuits. In the first proof-of-concept (Figure 7 a), we characterized the voltage output upon illumination, simultaneous illumination and dripping, and only dripping, employing a bridge rectifier for the D-TENG to convert from alternating current (AC) to direct current (DC) and an oscilloscope connected following the schematic circuit in Figure 7 b). In addition to the rectifier bridge, the PSC has been connected to a diode to prevent back current. In this way, both contributions are positive, and it is visible how the triboelectric AC voltage superimposes on the DC solar cell output (Figure 7 a).

In the second configuration, the energy management circuit includes a capacitor (either 2 µF or 100 nF). The output voltage can be acquired for the combined system or independently for the D-TENG. The schematic of the circuit is presented in Figure 7 c). This combination allows for the sequential and simultaneous actuation of the PSC and the D-TENG in a self-charging device. The comparison of Figure 7 d) between single sources allows for a rapid evaluation of both systems. As expected, the capacitor is charged faster by the solar cell, reaching its open voltage characteristic values (0.45 V for 0.5 suns) in less than 30 seconds, while the D-TENG reaches a saturation voltage of 125 mV in ca. 60 seconds.

The simultaneous activation of the PSC and D-TENG is presented in panels e) and f) for the energy management circuit included in panel c). The 2 µF capacitor reaches its maximum voltage rapidly when harvesting energy from both solar simulator light (0.5 suns) and Milli-Q droplets impacting at a frequency of 3 Hz. Employing the 100 nF capacitor provides an even clearer contribution from the dripping to the charging voltage (see Figure 7 f)). Thus, each drop impact event contributes with a sharp voltage increase of ~ 30 mV (see inset) with an envelope of ca. 150 mV, also consistent with the results presented in panel 7 d). These results demonstrate the synergistic combination of both energy harvesting devices, which can contribute to energy storage separately and simultaneously. Please note that a remnant contribution is observed in all these experiments after switching off the solar simulator, with maximum values of up to 100 mV. This effect is attributed to charge accumulation after illumination, which can persist for extended periods due to the capacitive contribution of the solar cell, especially under very low light illumination conditions. Dyakonov et al. [98]. demonstrated that perovskite solar cells can maintain open-circuit voltages even under low-light conditions, such as those found in moonlight. Furthermore, they reported a prolonged decay of the open-circuit voltage in the dark, with the voltage dropping to zero as long as 2 hours after the light was switched off. This slow decay indicates that charges remain stored at the electrodes for extended periods in the absence of light due to reduced recombination. In the case of our perovskite solar cells, a remnant voltage under dark conditions after illumination can also be explained by charge accumulation and a slow recombination of charges at the electrodes. The capacitive contribution of the trapped charges contributes to the overall voltage. In the absence of significant shunt pathways, these charges decay slowly, resulting in a persistent voltage even after illumination has ceased, which is enhanced due to the residual visible, NIR, or IR illumination.

In the third configuration, we have experimentally confirmed (Figure 7 g)) that the system can continuously power a red LED array, with a safety margin, using a custom-designed boost converter operating to step up the output voltage to the necessary driving voltage, nominally 1.8 V (8 mW). Also, the system provides sufficient power to intermittently switch on a green LED array of 2.5 V. The test setup is based on a discrete implementation on a protoboard (Figure 7 g)-left and g)-center), using an external pulse generator to control the switch actuation in an open-loop configuration (see photograph at Figure 7 g)-right) and **Video S3**). This configuration regulates the output voltage to drive the red LED at 1.8 V with the required 1 mA current for continuous illumination. The use of external control does not impact the feasibility analysis, as the system parameters fall within the operating range of existing commercial solutions based on closed-loop integrated switching regulators. Thus, for instance, the ADP5091 from Analog Devices supports input voltages as low as 0.38 V and outputs up to 5 V, while the BQ25504 from Texas Ins. operates from 0.1 V and boosts up to 5.5 V, both capable of handling input currents as low as 1 µA, and with an efficiency above 80 % for target use. Since the hybrid PSC/D-TENG provides peak powers of more than 10 mW, these power levels are well-suited within commercially available devices, further confirming the feasibility of operating LEDs under self-powered and autonomous conditions.

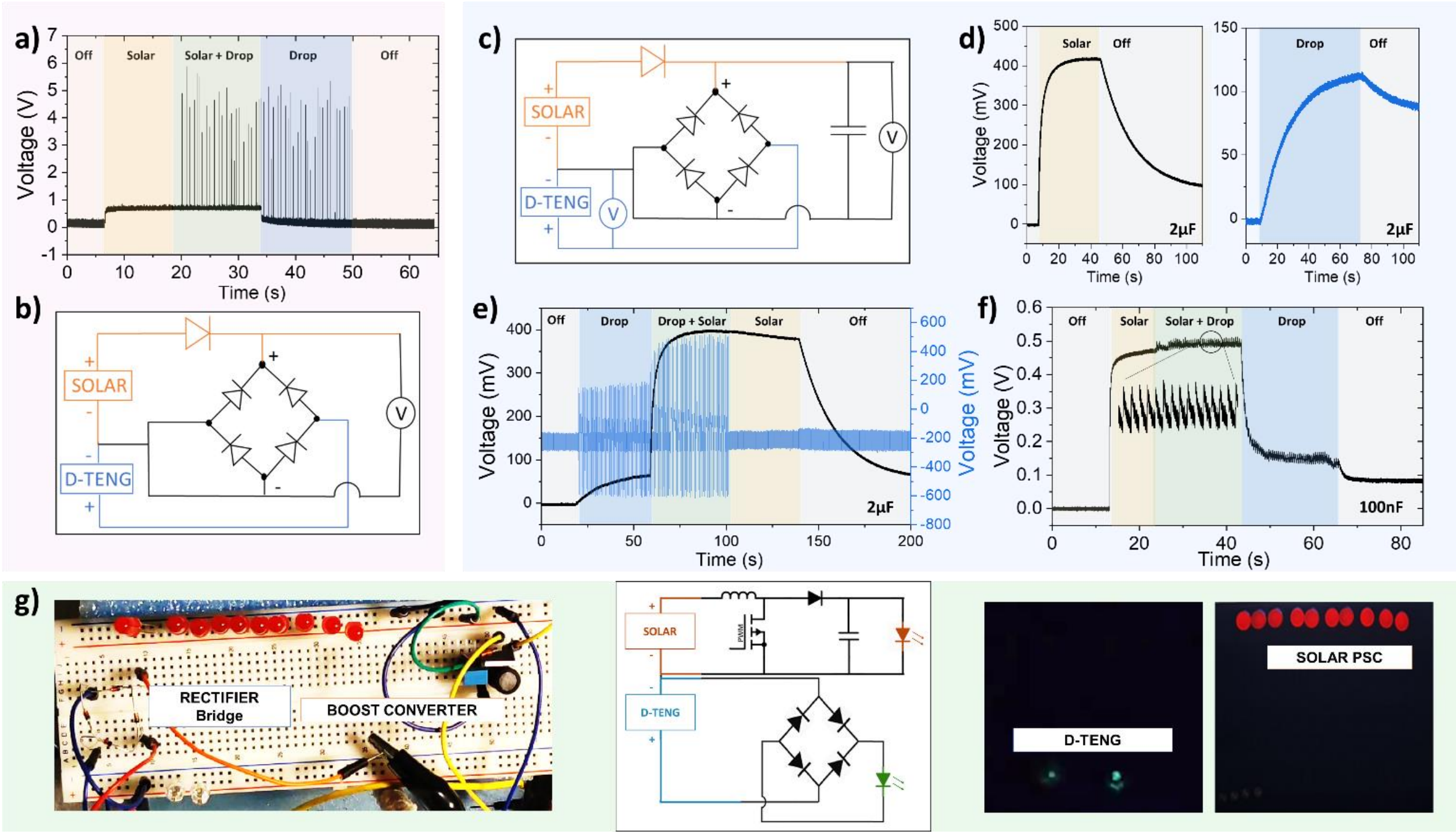

**Figure 7. Proof-of-concept of multisource energy harvesting.** a) Voltage output characteristics under various combinations of energy inputs for the hybrid PSC/D-TENG device. b) Energy management circuit used in panel (a), incorporating a full-wave bridge rectifier for AC-to-DC conversion of the triboelectric output and a diode to prevent back current from the PSC. c) Circuit configuration used for the self-charging demonstrations shown in panels (d–f). d) Charging of a 2 µF capacitor using either the PSC or the D-TENG, as indicated. e, f) Operation of the self-charging system with different combinations of dripping and illumination, using capacitors of 2 µF and 100 nF, respectively. g) Proof-of-concept device employing a boost converter to continuously power a red LED array via the PSC, and to enable instantaneous activation of green LEDs via the D-TENG. All experiments were conducted using a single PSC/D-TENG hybrid device with an FTO mutual electrode and equivalent 130 nm-$CF_x$ triboelectric and encapsulating layers. Photovoltaic parameters of the PSC (PCE = 15.7%) are provided in Table 2. All tests were performed under 0.5 sun illumination with MilliQ water droplets dripping at a frequency of 3 Hz. See also Video S3 in the SI.

Table S3 presents an overview of the reported combinations of solar cells and D-TENGs to contextualize these results. Unfortunately, the simultaneous harvesting mode is not demonstrated in most of these examples, and there is no standardized characterization of the D-TENG counterpart, which hampers the fair comparison and benchmarking of the results. The triboelectric output presented in this work is competitive with the values in the literature except when compared with top-bottom arrays,[5] thicker fluoropolymer counterparts, or complex interface engineering processes.[7,61] As mentioned above, the most popular hybrid system involves c-Si as the standardized photovoltaic counterpart, with only one additional example involving metal halide perovskite cells[61] (note that references [62–68,70] involve only solid-solid triboelectrification). The results obtained for simultaneous multi-source harvesting are enormously appealing, as the latest trends for harvesting low-intensity and diffuse light highlight the advantages of exploiting PSC technologies under illumination conditions compatible with cloudy and rainy days.[99] To the best of our knowledge (see Table S3), this is the first demonstration of the reliable development and durability of a hybrid thin film D-TENG and PSC.

## Conclusions

We have demonstrated the compatibility of the plasma-assisted fabrication of fluorinated thin film encapsulants with metal halide perovskite solar cells. The PSCs resisted the vacuum and energy conditions used to deposit $CF_x$ with thicknesses of tens to hundreds of nanometers on both Au and FTO/glass surfaces. Thus, the conformal growth of these fluorinated thin layers preserves the optoelectronic properties and extends the PSCs stability under simulated stress. The fluorine functional groups incorporated into the surface increase the surface hydrophobicity, protecting against vapor and liquid water interactions. The $CF_x$ layer also enables the combination with commercial epoxy encapsulants, improving stability even in underwater environments, and has previously demonstrated a high resistance to rain-erosion tests. The triboelectric performance of these Teflon-like thin films was tested with different types of water, such as Milli-Q and rain, in a switch electrode configuration, both in single- and array layouts. A hybrid perovskite and D-TENG device has been assembled and tested. The combined hybrid harvester operated for more than 5 hours under continuous illumination at 0.5 suns and periodic Milli-Q water dripping. We have also demonstrated a proof-of-concept of the synergistic combination of the two energy-harvesting mechanisms. The energy management circuit has been designed to favour the positive contribution of both cells and tested in a self-charging configuration. Both PSC and D-TENG have been employed to charge a supercapacitor, with the dynamic being much faster for the solar cell, even under low illumination conditions. We have also experimentally confirmed that the system can continuously power a red LED array using a custom-designed boost converter. The fabrication of the Teflon-like thin film is carried out at room temperature, under mild vacuum and plasma power conditions, and it is fully compatible with current interface engineering approaches for enhancing the output power of triboelectric nanogenerators and with the different architectures and compositions of the PSCs. We trust these results will boost the development of plasma and vacuum-phase encapsulants and pave the way for the hybridization of perovskite solar cell devices with triboelectric nanogenerators.

### Experimental Section

*Materials.* To synthesize perovskite solar cells (RbCsMAFA), FTO (fluorine-doped tin oxide Pilkington TEC 15, resistance 15/square, 82-84.5% transmittance) substrates coated on glass were purchased from XOP Glass. Both titanium diisopropoxide bis(acetylacetonate) (75% in 2-propanol) and $TiO_2$ paste (18NRT) were purchased from Sigma-Aldrich meanwhile absolute ethanol (99.9%) was commercialized by Scharlau. Solvents (N,N-dimethylformamide (DMF), dimethyl sulfoxide (DMSO), chlorobenzene (ClBn) and acetonitrile(AN)) were purchased from Acros Organics. Formamidine hydroiodide (FAI, >98%), methylamine hydroiodide (MAI, >99.0%), lead (II) iodide ($PbI_2$, 98.0%), and lead (II) bromide ($PbBr_2$) both for perovskites were supplied by TCI. Caesium iodide (CsI, 99.9%) and Rubidium iodide (RbI, 99.9%) were acquired from Alfa Aesar and Sigma-Aldrich respectively. Furthermore, 2,2,7,7-tetrakis[N,N-di(4-methoxyphenyl)amino]-9,9-spirobifluorene (Spiro-OMeTAD, >99.9% commercialized under the name of SHT-263 Solarpur), lithium bis(trifluoromethanesulfonyl)imide (LiTFSI, 99.95%), tris(2-(1H-pyrazol-1-yl)-4-tert-butylpyridine)-cobalt(III)tris(bis(trifluoromethylsulfonyl)imide), commercially known as FK209 Co(III) (98%) and 4-tert-butylpiridine (98%), were purchased from Sigma-Aldrich. Finally, gas octafluorcyclobutane ($C_4F_8$, 99.997%) commercialized by Linde has been used as the fluorinated precursor for $CF_x$ coatings, and the UV light-curable E132 epoxy suitable for solar cell encapsulation was purchased from Ossila.

*Perovskite Solar Cells fabrication.* Perovskite solar cells (PSCs) were fabricated following the standard procedures detailed in **Supporting Information S15** according to [19,22,28].

*$CF_x$ encapsulant thin film deposition.* Fluorinated polymeric thin layers ($CF_x$) have been fabricated using Plasma Enhanced Chemical Vapor Deposition (PECVD) in a parallel plate capacitive radiofrequency reactor on the PSCs as well as onto silicon and fused silica references [72,73] at a base pressure of $5x10^{-2}$ mbar. Then a mixture of 50%

Ar + 50% of perfluorocyclobutane ($C_4F_8$) at 0.1 mbar as working pressure was introduced in the plasma reactor. Mild RF power conditions were applied to the bottom electrode acting as sample holder, with a fixed self-induced negative bias voltage of 50 V (from 15 to 20 W approximately), while the top electrode was grounded. Thicknesses of the deposited $CF_x$ layers have been controlled by the deposition time (from 5 to 30 min). Control over the fluorinated functional species present in the $CF_x$ layer can be optimized for triboelectric purposes working, for example, with a fixed applied power of 20 W (maximum value reached during the fabrication of the $CF_x$ encapsulant), which involved a self-adjustment of the autobias at 17 V for a glass substrate. This leads to lower ionic energy affecting less breaking of fluorinated fragments and, in consequence, preserving better the Teflon-like polymeric character instead of promoting crosslinking.

*Epoxy resin-based encapsulation.* Small pieces of a glass coverslip of 2 cm x 1.5 cm are cut and rinsed with acetone and isopropanol cycles. A single drop of the E132 Ossila epoxy (3,4-epoxycyclohexylcarboxylate) is dispensed by a pipette onto the top surface of the perovskite solar cell (top electrodes side), and the glass cover is placed above, pressing slightly to remove any possible bubble. Finally, curing can be achieved by exposition to a low-intensity UV lamp for 20 min following the seller's instructions.

*Assembly of the hybrid PSC / D-TENG harvester.* After thoroughly cleaning the FTO/glass substrate, we proceed to the deposition of the perovskite solar cell and the subsequent encapsulation with the $CF_x$ hydrophobic layer. In the reverse of the cell (glass side), we deposit another $CF_x$ coating acting as a triboelectric layer so that the multisource device becomes fully encapsulated with the thin $CF_x$ film. We deposit the top metallic (5nm of Ti/90nm Au) electrode of the D-TENG by thermal evaporation using masks of "L" shape according to [56]. The hybrid devices also incorporate the Ossila commercial epoxy resin-based encapsulant on the PSC side following the supplier instructions: press with glass cover-slip of about 2 cm x 1.5 cm and cure with ultraviolet light. Finally, we connected with copper wires, conductive silver epoxy, and a commercial sealant that provides robustness. The common electrode of the PSC deposited directly on the FTO will also serve as the bottom electrode of the D-TENG device.

*Characterization techniques.* Microstructure analysis was carried out by Scanning Electronic Microscopy (SEM) using an SEM Hitachi 4800 (2 kV y 10 mA) in secondary and backscattered electron modes to determine the thickness and uniformity of the layers in cross-section. The chemical composition of the surfaces was studied by X-ray Photoemission Spectroscopy (XPS) with a SPECS XRC 1000 analyzer using the non-monochromatic line of Al Kα working in a constant energy mode of 20 eV for the survey and 10 eV for high-resolution binding energy (BE) zones. The binding energy scale has been calibrated with the C-C(H) photopeak assigned at 284.5 eV. Optical characterization has been performed by Variable Angle Spectroscopic Ellipsometry (VASE) to obtain optical thickness and refractive index in function of the polarity changes when the light is reflected. The Cauchy model is used to fit the experimental data. Ultraviolet-visible (UV-Vis-NIR) Spectrophotometry was performed with a PerkinElmer Lambda 750 UV/vis/NIR in the range of 200-2200 nm. Nanoindentation tests using the Continuous Stiffness Measurement (CSM) technique with a maximum load of 1mN have been performed at room temperature in KLA G200x Nanoindenter for hardness and elastic modulus analysis. A diamond Berkovich tip has been employed over twenty-five measurements. A thicker $CF_x$ coating of 700 nm has been used to estimate the mechanical properties at the 10% of the total thickness to avoid the substrate influence.

Wetting properties have been characterized by static Water Contact Angle (WCA) in the OCA20 goniometer from DataPhysics equipment. 2 µL droplets of Milli-Q water were deposited onto the surface considering a statistical analysis over 5 repetitions for averaged values with a margin of error below 3% of the hydrophobic / hydrophilic surface behaviour.

Photovoltaics parameters have been extracted from the current density-voltage (J-V) curves of the PSCs which were recorded under a solar simulator (ABET-Sun2000) with AM1.5G filter at 100 $mW{\cdot}cm^2$. Photovoltaic parameter statistics are evaluated in terms of number of samples and frequency for pristine (reference) and surface coated (before and after) devices under 1 sun illumination in reverse scan with a scan rate of 200 $mV·s^{-1}$ from 1.2 to -0.1 V for each electrode. Samples were illuminated through a black mask with an aperture of 0.14 $cm^{-2}$ to calculate the photovoltaic parameters. From these curves, it is possible to extract relevant photovoltaic parameters such as open circuit voltage ($V_{oc}$), short-circuit current density ($J_{sc}$), maximum power ($P_{max}$), Fill Factor (FF), and Power Conversion Efficiency (PCE). Normalized PCE was also determined and plotted in the graphs to establish the comparison between different PSCs. The photovoltaic parameters of encapsulated and pristine (before encapsulation) perovskite solar cells have been studied based on a statistical analysis of variance using ANOVA[77] on a total of more than 40 electrodes or half-cells.

Long-term stability characterization tests were performed under different environments to determine the PSC behavior while working in extreme conditions in an ad-hoc automatized system that permits monitoring the full J-V curves sequentially and statistical analysis of the cell behavior. ABA LED Solar Simulator from Newport (Model LSH-7320, AM1.5G) calibrated for 1 sun illumination has been employed. The samples were placed inside a homemade hermetic container allowing their illumination through a transparent window. Continuous gas flows (up to 120 sccm) of nitrogen ($N_2$), oxygen ($O_2$), or air ($O_2$/ $N_2$) were controlled by a rotameter. The relative humidity is controlled by the relative flows of the dry and the moisture saturated (using a water bubbler) gases. A relative humidity sensor is placed in the container outlet to record the actual degree of moisture exposition. PSCs have been masked to illuminate an effective area (0.14 $cm^2$) to determine the correct photovoltaic parameters, whereas the electric contact of the top electrodes has been reinforced with silver conductive paint. The dry stability test was performed for 60 h under 1 sun illumination, whereas the 100 % RH test was kept. Taking advantage of the controllable temperature of the sample holder, a third aggressive experiment was performed, heating up to 50 ºC in addition to a relative humidity of 100 %. Additionally, long-term stability test over encapsulated epoxy/$CF_x$/PSC and $CF_x$/PSC was perfomed in the same homemade setup for more than 300h under $N_2$ flow and 50%RH at room temperature. Those samples were also tested under harsh environmental conditions with simultaneous illumination: high humidity (80%RH) of air flow, with changing temperature from RT for 60h to 100ºC for 170h after. Environmental stress-simulated cycling was performed placing samples into the universal testing materials machine (Signeblock: a9fe07d986f95ae20575d9bf1ad2665ca4b8074e7b65df6dd8d4bbe33e9b8d51) with cycles of 20% RH and 50ºC for 1 hour followed by 80%RH and 10ºC for 1 hour. Photovoltaic parameter measurements after each cycle have been carried out with the already mentioned Solar Simulator. It should be noted that the stability tests reported here were initiated more than two years ago, using devices representative of our laboratory performance at that time (average PCE ≈ 13–14%). While these values are below current benchmarks, current higher-efficiency cells (also synthesized currently in our laboratory) generally exhibit equal or greater intrinsic stability. Therefore, the benefits of $CF_x$-based encapsulation demonstrated here are expected to be even more pronounced in future work with state-of-the-art devices.

Two different parameters have been introduced to characterize the stability behavior of the PSC ($T_{80}$ and $T_{s80}$))[79] $T_{80}$ is defined as the time in which the PCE (Normalized PCE) decays up to 80 % of its initial value while the lifetime parameter $T_{s80}$ measures also the decay of its 80 % of efficiency without considering the “burn-in period”, this is the initial fast exponential decay of the cell. Thus, $T_{s80}$ is always longer than $T_{80}$.
Finally, a liquid water immersion test has been performed by assembling electrical connections on the top and bottom PSC electrodes directly to the power supply. PSCs placed in a transparent cuvette filled with milli-Q water were exposed to the solar simulator's light source (rotated for horizontal illumination) calibrated for 1 sun (AM1.5G).

For the solid-solid durability test, a magnetic shaker (Smart Shaker K2007E01 from The Modal Shop) attached to a force sensor (IEPE model 1053V2 from Dytran Instrument, Inc.) was used for the tapping movement. This magnetic shaker can vary the frequency and the force of the tapping. The response of the devices was measured by a Keithley 2635A.

*D-TENG harvester characterization.* For the triboelectric characterization, 35 µL droplets of different nature (Milli-Q and rainwater, with electrical conductivities of 6 µS/cm and 60 µS/cm, respectively) were dispensed from a syringe with a 0.25 mm diameter grounded (or not) metal tip connected to a peristaltic pump (which allows to control the dripping frequency). Rainwater was collected on the 8$^{th}$ of February 2024 in Seville, Spain. These drops are dispensed from a base with an adjustable height and fall on the sample placed in a holder with an adjustable tilting angle. The output voltage of the D-TENG is measured using a PicoScope 3000D Series oscilloscope. Measurements were performed in a lab environment with humidity of 65 %, and 20 º C. Water conductivity measurements were performed using a conductivity probe 731-ISM Mettler Toledo InLab as the conductivity module for the Mettler Toledo SevenExcellence multiparameter. Short-circuit current and charge transferred were acquired using a Keithley 2635A. The load impedances were changed using a resistance box that allows sweeping from 100 kΩ to 100 MΩ in parallel with the oscilloscope probe resistance of 10 MΩ. For the statistical evaluation of the D-TENG performance, we analyzed several output parameters, including mean and maximum power, energy per cycle (see Table S2 for details), maximum voltage in the positive and negative branches (V+/V-), and peak shape characterized by the full width at half maximum (FWHM+/FWHM-). The statistical treatment was carried out using the open-access software NanoDataLyzer, considering a minimum of 10 events or drops per resistance.[103]

*Characterization of the hybrid PSC / D-TENG harvester.* For the hybrid characterization, the holder was tilted 45º so that the LED of the solar simulator shone perpendicular to the sample with a power of 0.5 suns. At the same time, the electrical connections were connected to the photovoltaic parameter measurement system using the Keithley sourcemeter. Measurements were continuously recorded, and long cell stability tests were performed without the illumination mask to avoid water accumulation issues. Simultaneously, the D-TENG system was characterized following the procedure previously explained in the D-TENG harvester characterization section. For the simultaneous measurement, both tests were launched simultaneously using a full rectifier bridge circuit (**Figure 6b** and **Figure 7**)) to check the veracity of the multisource hybrid system. To minimize power losses, we selected ultra-low-leakage diodes from the commercial BAS45A family, which exhibit an inverse saturation current on the order of 1 pA. Charge-capacitor tests (Figures 7d–f) were conducted using the same circuit configuration, with capacitors of different values (2 µF and 100 nF). Finally, we experimentally confirmed that the system can continuously power a red LED array with a safety margin by employing a custom-designed boost converter that steps up the output voltage to the required driving level (nominally 1.8 V). Additionally, the TENG output provides sufficient power to intermittently switch on a green LED array rated at 2.5 V. The test setup was implemented on a protoboard (Figures 7g-left and 7g-center) using discrete components and an external pulse generator to control the switching in an open-loop configuration (see photograph in Figure 7-left), demonstrating the feasibility of our hybrid D-TENG/PSC approach.

**Conflicts of interest**

There are no conflicts to declare.

**Data availability**

Data availability is accessible by direct request to the authors.

**Credit authorship**

FNG, XGC, LCB, CLS, AD, JMOP, JCS, and JPE conducted fabrication and characterization experiments. XGC and FNG were responsible for the software development. JRSV and XGC developed the automated system for solar cell stability measurements. AG and GL assisted in the implementation of the power management system for hybrid solar - rain cells. JCSL performed the mechanical characterization of the CF x thin f ilms. CLS, AnaB, AB, and JRSV did the conceptualization. CLS, AnaB, and JRSV wrote the original draft of the article. All the authors corrected and contributed to the final version of the manuscript

**Acknowledgments**

The authors thank the projects PID2022-143120OB-I00, PCI2024-153451 and TED2021-130916B-I00 funded by MCIN/AEI/10.13039/501100011033 and by "ERDF (FEDER)" A way of making Europe, Fondos NextgenerationEU and Plan de Recuperación, Transformación y Resiliencia. Project ANGSTROM was selected in the Joint Transnational Call 2023 of M-ERA.NET 3, an EU-funded network of about 49 funding organizations (Horizon 2020 grant agreement No 958174). XGC acknowledges the FPU program under the grant number FPU19/01864. FNG and CLS acknowledges the "VII Plan Propio de Investigación y Transferencia" of the Universidad de Sevilla. The project leading to this article has received funding from the EU H2020 program under grant agreement 851929 (ERC Starting Grant 3DScavengers).

# SUPPORTING INFORMATION

## Water-resistant hybrid perovskite solar cell - drop triboelectric energy harvester

Fernando Núñez-Gálvez,[a,b] Xabier García-Casas,[a] Lidia Contreras Bernal,[a,c] Alejandro Descalzo,[a] José Manuel Obrero-Pérez,[a] Javier Castillo-Seoane,[a] Antonio Ginés,[d] Gildas Leger,[d] Juan Carlos Sanchez-Lopez,[e] Juan Pedro Espinós,[a] Ángel Barranco,[a] Ana Borrás,[a]* Juan Ramón Sánchez-Valencia,[a]* Carmen López-Santos[a,b]*

f) Nanotechnology on Surfaces and Plasma Laboratory, Consejo Superior de Investigaciones Científicas (CSIC), Materials Science Institute of Seville (CSIC-US). c/ Américo Vespucio 49, 41092, Seville (Spain).
g) Departamento de Física Aplicada I, Universidad de Sevilla, c/ Virgen de Africa 7, 41011, Seville (Spain).
h) Dpto. Química Física. Facultad de Química. Universidad de Sevilla. C/Profesor García González 41012, Seville (Spain).
i) Instituto de Microelectrónica de Sevilla (IMSE-CNM), Universidad de Sevilla – CSIC, Avda. Americo Vespucio 28, 41092, Seville (Spain).
j) Materials Science Institute of Seville (CSIC-US). c/ Américo Vespucio 49, 41092, Seville (Spain).

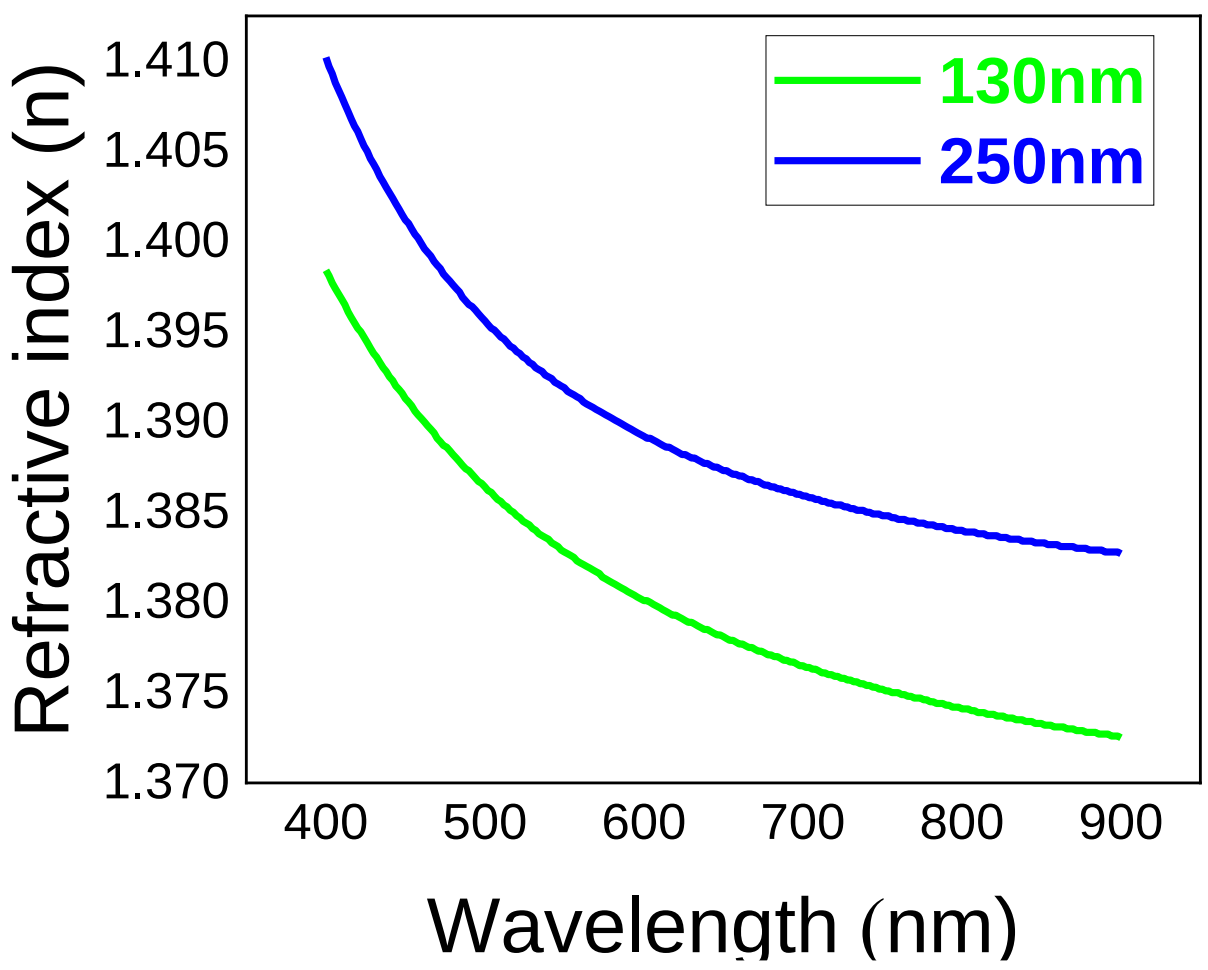

**Figure S1.** Refractive index dispersion relation according to Cauchy model extracted from ellipsometry analysis of $CF_x$ coatings with different thickness values. Optical thickness values obtained by fitting Cauchy model for the ellipsometric spectral are in good correspondence with thickness measured by SEM.

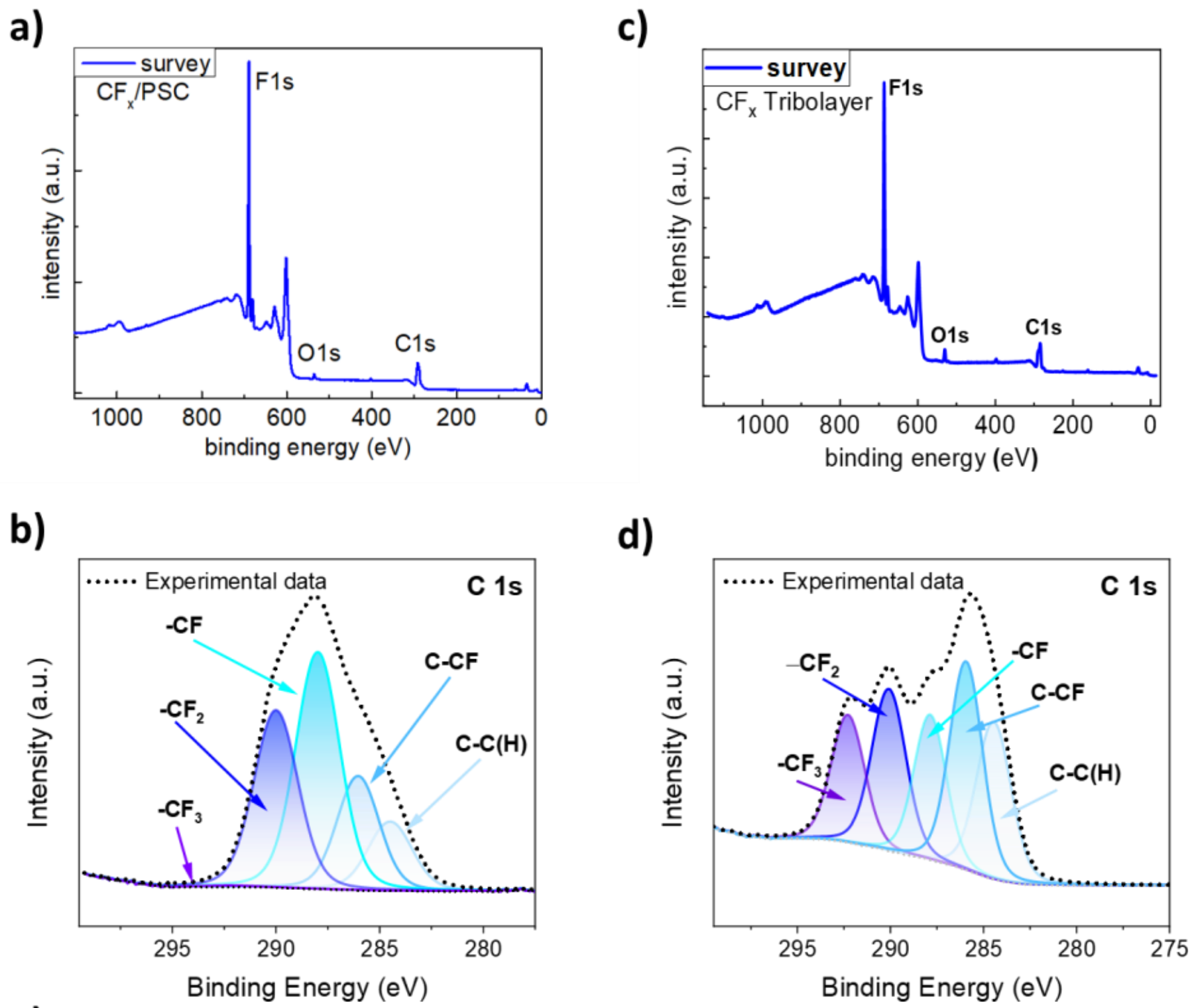

e)

| Atomic concentration (%) | C | O | F |
|---|---|---|---|
| CF$_x$/PSC | 47.3 | 2.2 | 50.5 |
| CF$_x$ TL | 56.6 | 4.5 | 38.9 |

f)

| Atomic concentration (%) | C-C/C-H | C-CF | CF | CF$_2$ | CF$_3$ |
|---|---|---|---|---|---|
| CF$_x$/PSC | 11.6 | 18.4 | 39.9 | 29.9 | 0.2 |
| CF$_x$ TL | 20.4 | 25.7 | 17.5 | 20.6 | 15.8 |

**Figure S2. XPS characterization of CF$_x$ layers**. CF$_x$ deposited over PSC for encapsulation (a,b) and as a tribolayer over glass reference substrate (c-d): a,c) survey spectrum and b,d) detailed fitting of the C1s photopeak region indicating the main fluorinated functional groups present at this surface. e) Table of atomic concentration (%) of both CF$_x$ surfaces and f) atomic percentages corresponding to the functional species detected from the C1s photopeak deconvolutions. The fluorine content of the tribolayer and functional group distribution differ markedly from the encapsulant layer. The CF$_x$ tribolayer shows an F/C ratio of 0.68 versus 1.07 for the encapsulant and is enriched in highly electronegative groups (-CF$_2$ and -CF$_3$). These groups account for 36.4% of the atomic composition, about 22% higher than the encapsulant, which is mainly rich in –CF and lacks -CF$_3$ species.

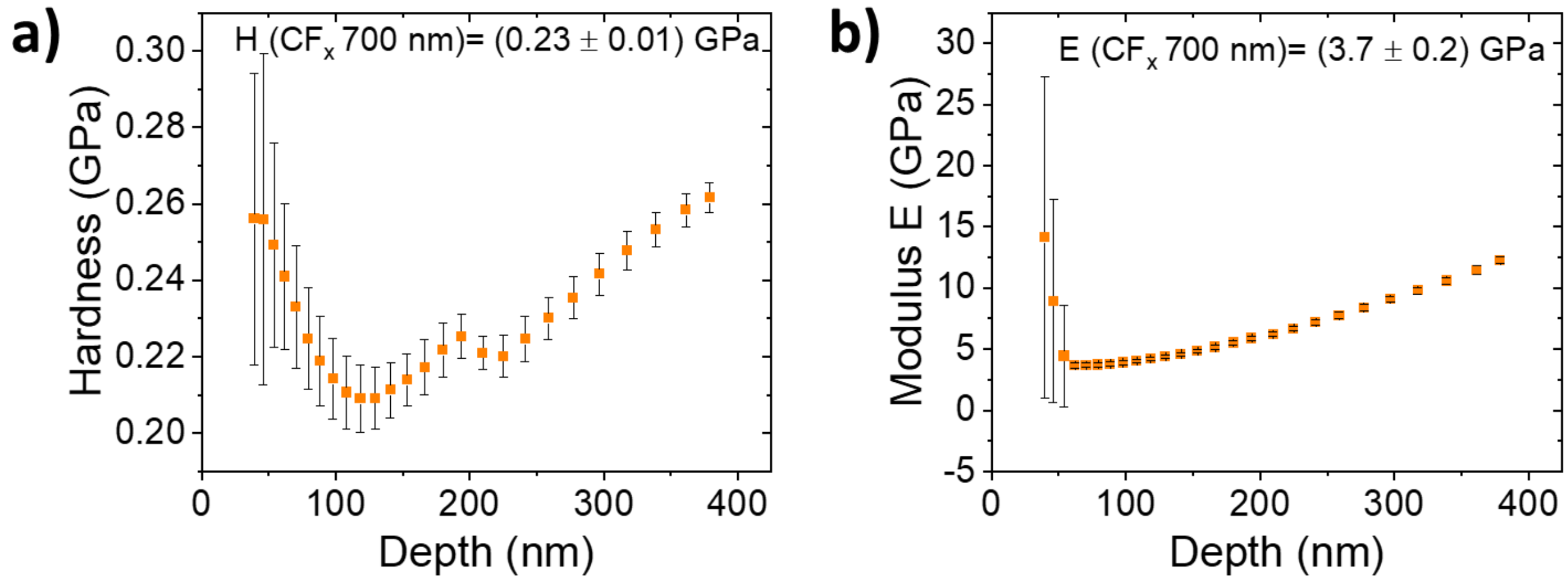

**Figure S3**. Nanoindentation tests by the Continuous Stiffness Measurement (CSM) method: hardness, H (a), and elastic modulus, E (b), as a function of penetration depth for a $CF_x$ layer deposited on a silicon wafer. Due to the technical limitations of nanoindentation, a 700 nm-thick film was used for the characterization.

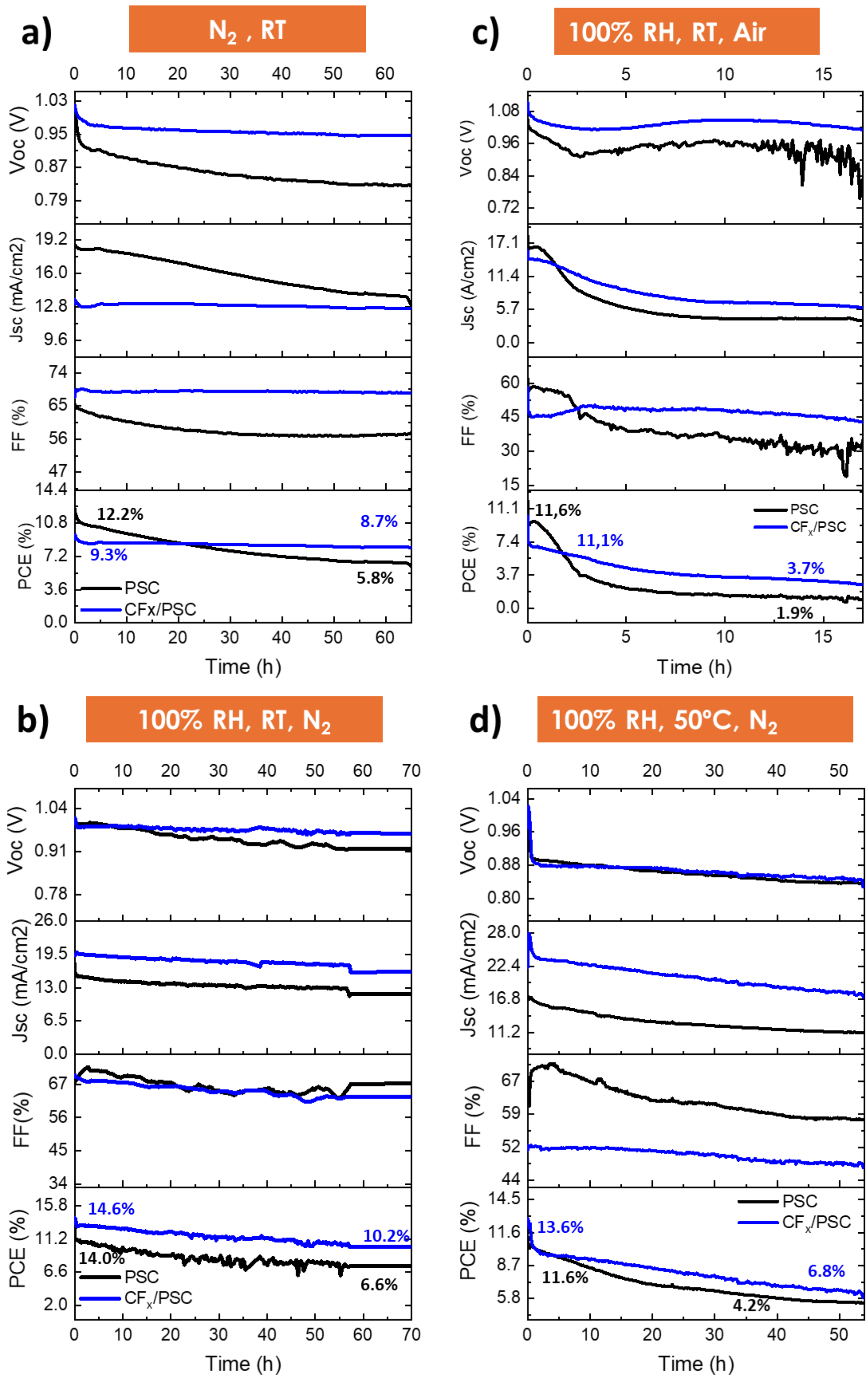

**Figure S4.** Complete photovoltaic parameters evolution extracted from the J-V curves comparing the as-prepared PSC and after encapsulation with the $CF_x$ (130 nm) under continuous illumination (1 sun, AM1.5G) at different stressing conditions: a) $N_2$ at room temperature (RT), b) $N_2$ at 100% relative humidity (RH) and RT, c) air at 100% RH and RT, d) $N_2$ at 100% relative humidity (RH) and 50ºC.

**Table S1.** Photovoltaic lifetime parameters ($T_{80}$ and $T_{s80}$) extracted from the stability tests.

| Stability test conditions | Sample | $T_{80}$ (h) | $T_{s80}$ (h) |
|---|---|---|---|
| $N_2$<br>RT | $CF_x$/PSC | >65 | >200 |
| | PSC | 8 | 27 |
| $N_2$<br>RT, 100% RH | $CF_x$/PSC | >30 | 60 |
| | PSC | 7 | 27 |
| Air ($O_2$/$N_2$)<br>RT, 100% RH | $CF_x$/PSC | 2.5 | 5 |
| | PSC | <0.5 | 2 |
| $N_2$<br>50ºC, 100%RH | $CF_x$/PSC | <10 | 40 |
| | PSC | 8 | 12 |

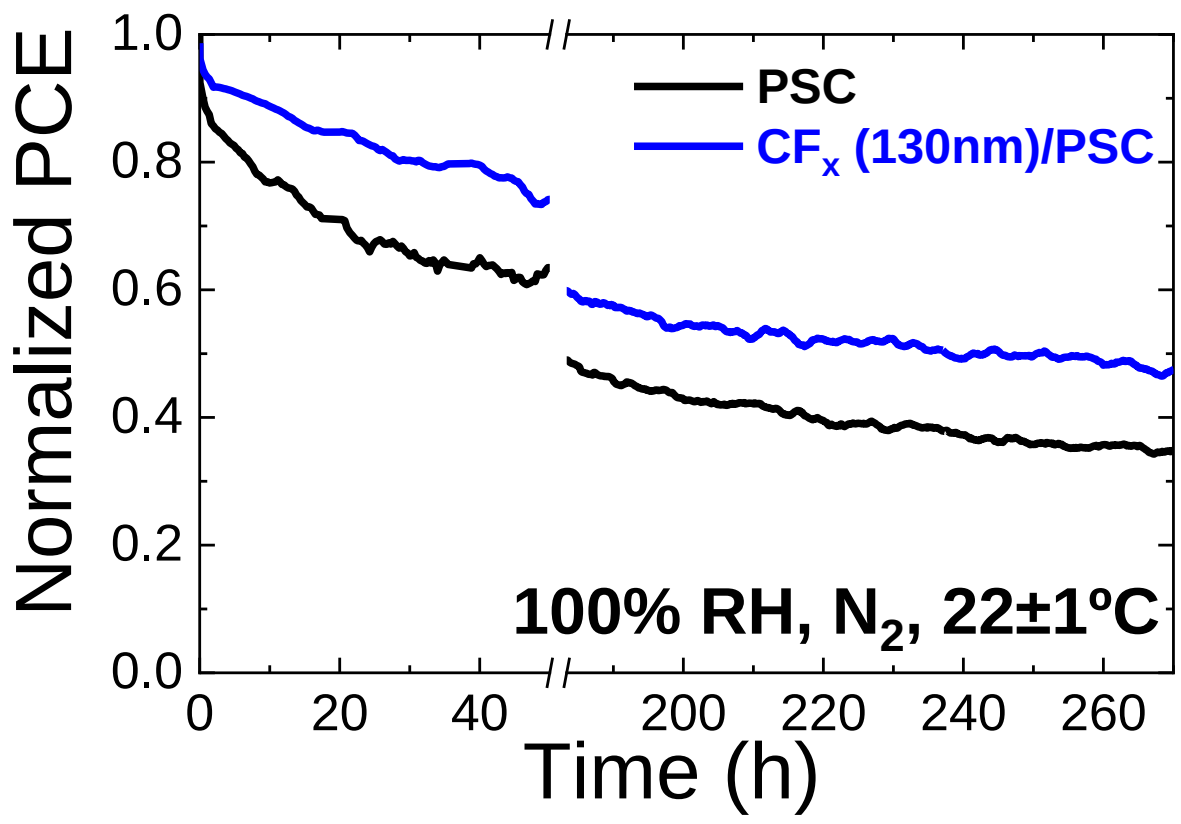

**Figure S5.** Normalized PCE evolution of the as-prepared PSC and after encapsulation with $CF_x$ (130 nm) under continuous illumination (1 sun, AM1.5G) and saturated humidity nitrogen flux at room temperature (22ºC).

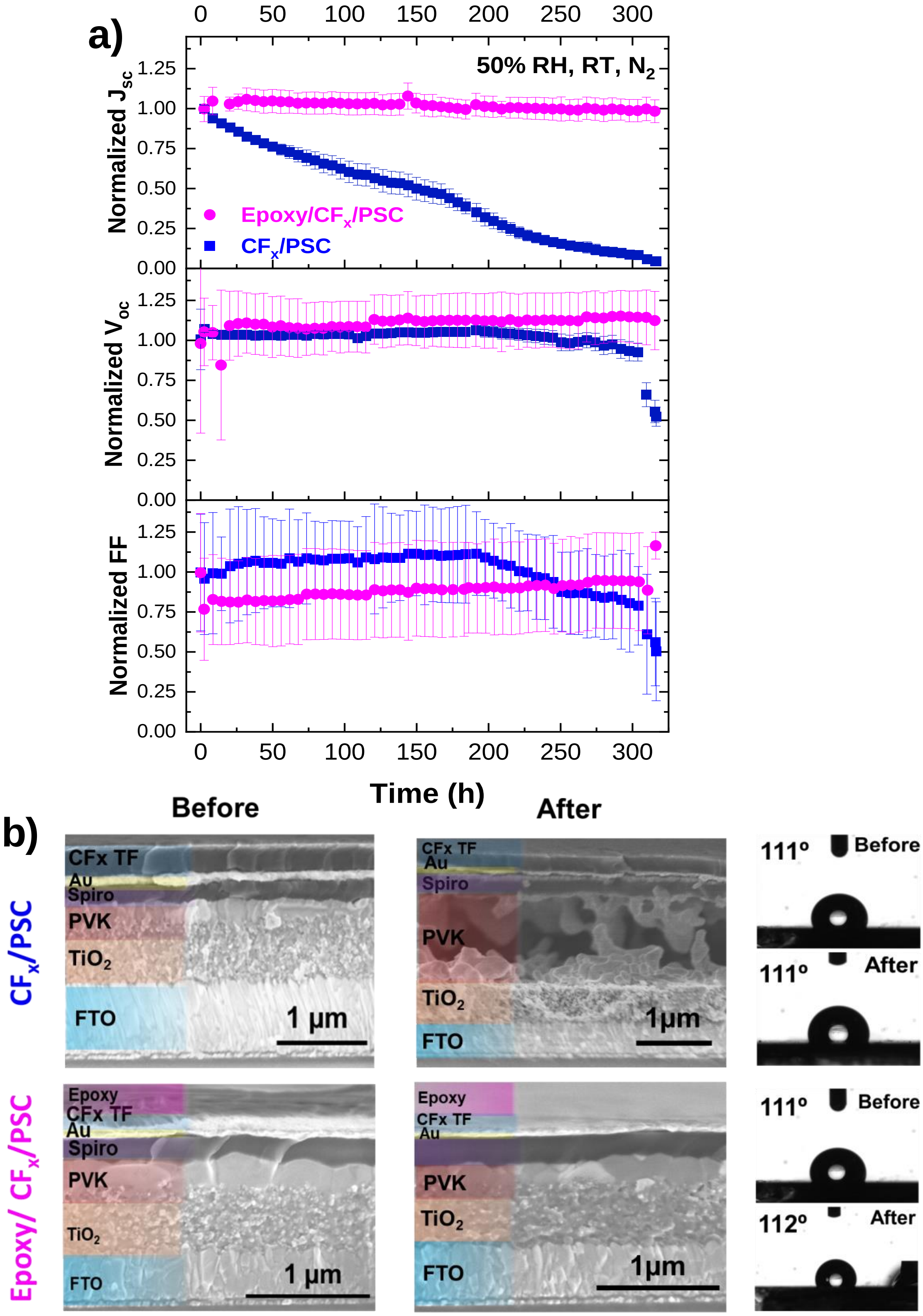

**Figure S6.** a) Complete photovoltaic parameters evolution ($V_{oc}$, $J_{sc}$ and FF) extracted from the J-V curves comparing the as prepared $CF_x$ (130 nm) encapsulated PSC with and without Epoxy encapsulation under continuous illumination (1 sun, AM1.5G) at $N_2$ at room temperature and 50% RH for more than 300 h testing. b) Cross-sectional SEM images of Epoxy/$CF_x$/PSC and $CF_x$/PSC devices and surface wetting characterization before and after the stability test.

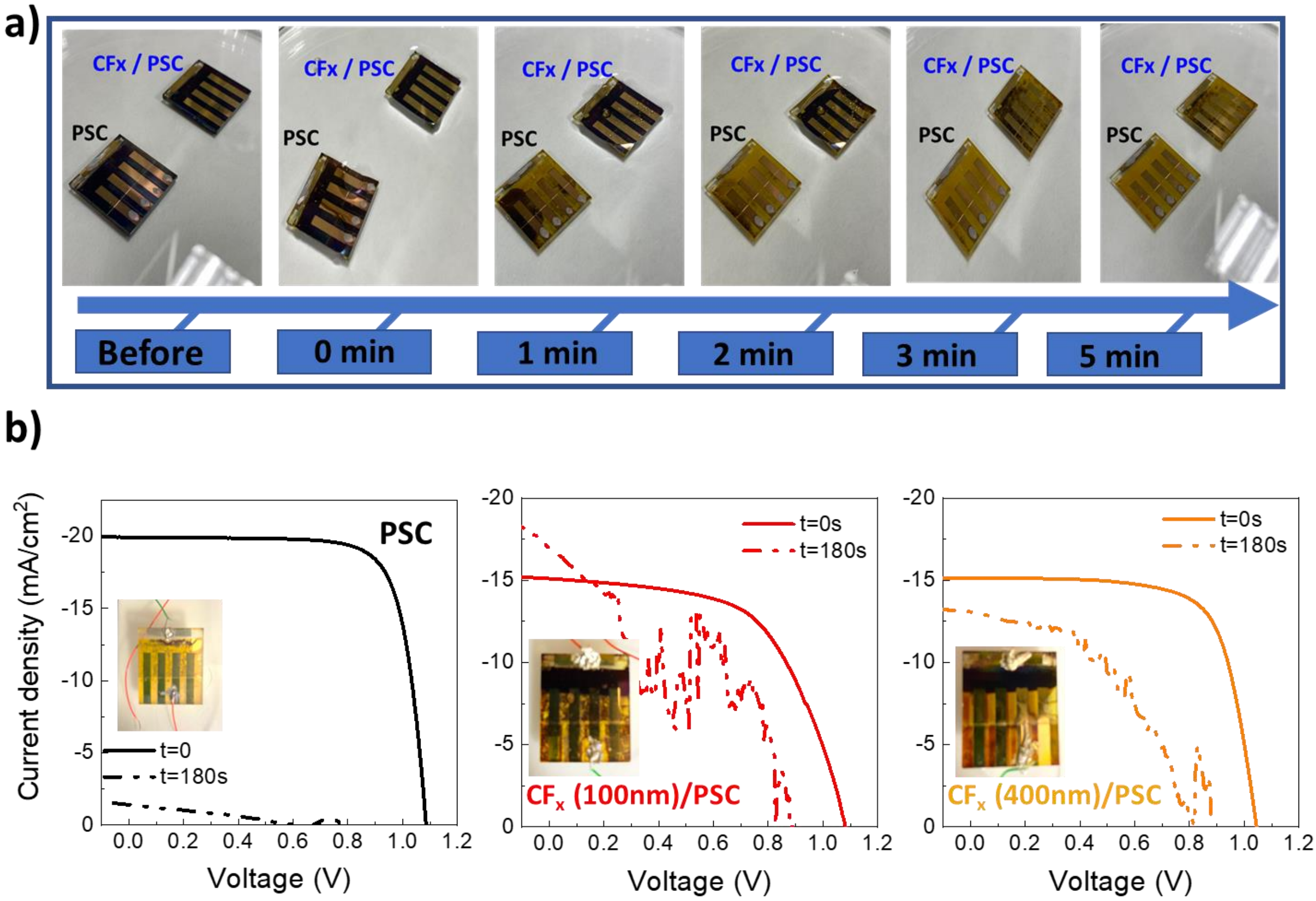

**Figure S7.** a) Liquid water immersion test of PSC reference and 100 nm $CF_x$ encapsulated PSC at increasing periods of time up to 5 min. b) Photocurrent density-voltage curves for reference and encapsulated devices before and after 180 s of water immersion.

**Table S2.** Drop-triboelectric energetic parameters corresponding to the $CF_x$ D-TENG with two water types, calculated from the drop harvesting test for a 0.15 $cm^2$ of active surface area (considering the drop-surface contact area by image analysis for a 35 µL of volume drop dripping from a 30 cm of height). Calculations were carried out from the V-t curves measures by using the $R_{load}$ that maximized the Power.

| | **Rainwater ($R_{Load}$=476 kΩ)** | **Milli-Q ($R_{Load}$=826 kΩ)** |
|---|---|---|
| **P** | 0.6 mW | 85.8 µW |
| **P/A** | 4 mW/$cm^2$<br>40 W/$m^2$ | 0.57 mW/$cm^2$<br>5.72 mW/$m^2$ |
| **E** | 33.8 nJ | 9.10 nJ |
| **E/A** | 225.3 nJ/$cm^2$<br>2.25 mJ/$m^2$ | 60.67 nJ/$cm^2$<br>0.61 mJ/$m^2$ |

The previous energetic parameters are calculated from the expressions:

$$Instantaneous\ peak\ Power:\ \ P = \frac{V_p^2}{R}$$

$$Peak\ power\ density:\ \ \frac{P}{A} = \frac{V_p^2}{R \cdot A}$$

$$Energy:\ \ E = \int_0^t \frac{V^2}{R}\, dt$$

$$Energy\ density:\ \ \frac{E}{A} = \frac{\int_0^t V^2 dt}{R \cdot A}$$

where V is the voltage and $V_p$ the maximum voltage peak in the V-t curves; A is the active surface area (A=0.15 $cm^2$), and R is the resistance used from the V-t measurements (indicated in the Table S2).

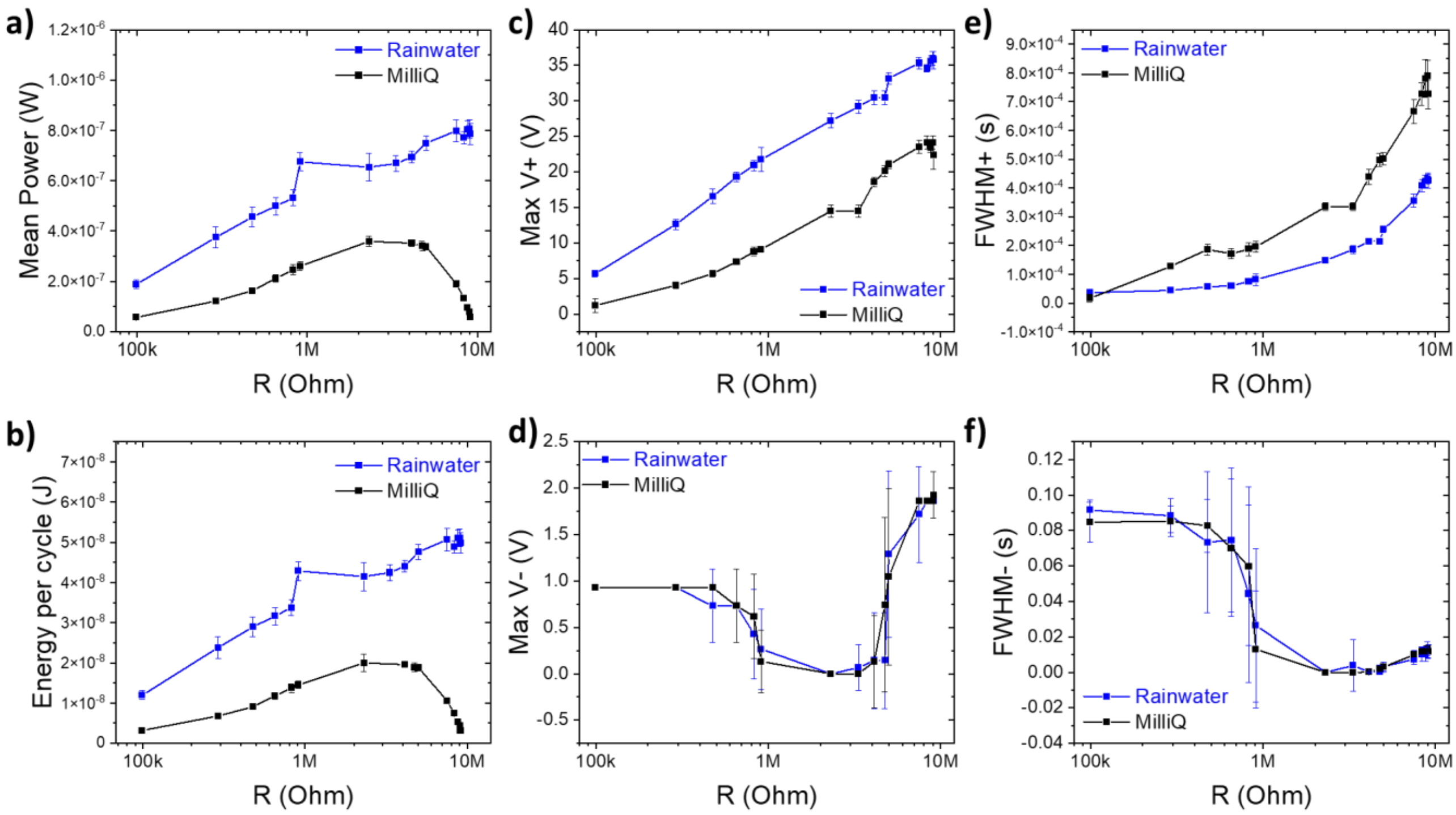

**Figure S8. Drop-triboelectric parameters of the $CF_x$ D-TENG as function of the electrical load:** a) mean power (as the full signal integrated area), where the maximum of the curve corresponds to the internal resistance of the oscilloscope probes (10 MΩ), hindering the characterization at higher resistances; b) energy per cycle and c-d) maximum V+/V- output peak power for increasing load resistances. e-f) Full Width at Half Maximum (FWHM+/FWHM-) of the positive and negative signals respectively, attending the increasing resistance value. Error bar values were obtained for a minimum of 10 drops of Milli-Q water for each point in the curve.

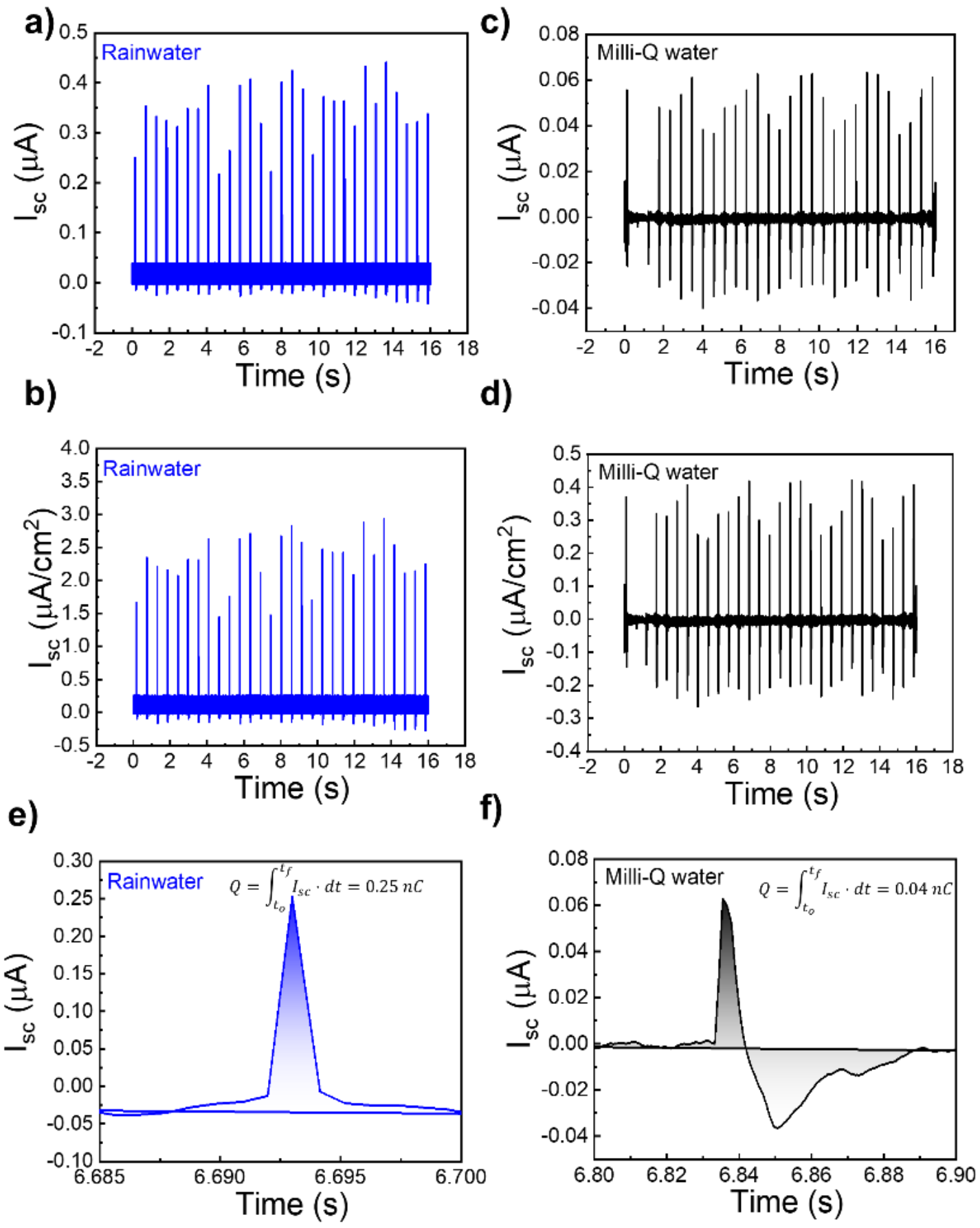

**Figure S9**. Short-circuit current evolution for an aged $CF_x$ D-TENG device for a-b) rainwater and c-d) Milli-Q water measured under conditions: 45º tilt , 2 Hz, ungrounded tip, 35 μL of drop volume and 30 cm of height. Note that b) and d) are divided by the active area of the drop-layer contact (0.15 cm$^2$). Charge-transferred calculation extracted from the peak for e) rainwater and f) Milli-Q water.

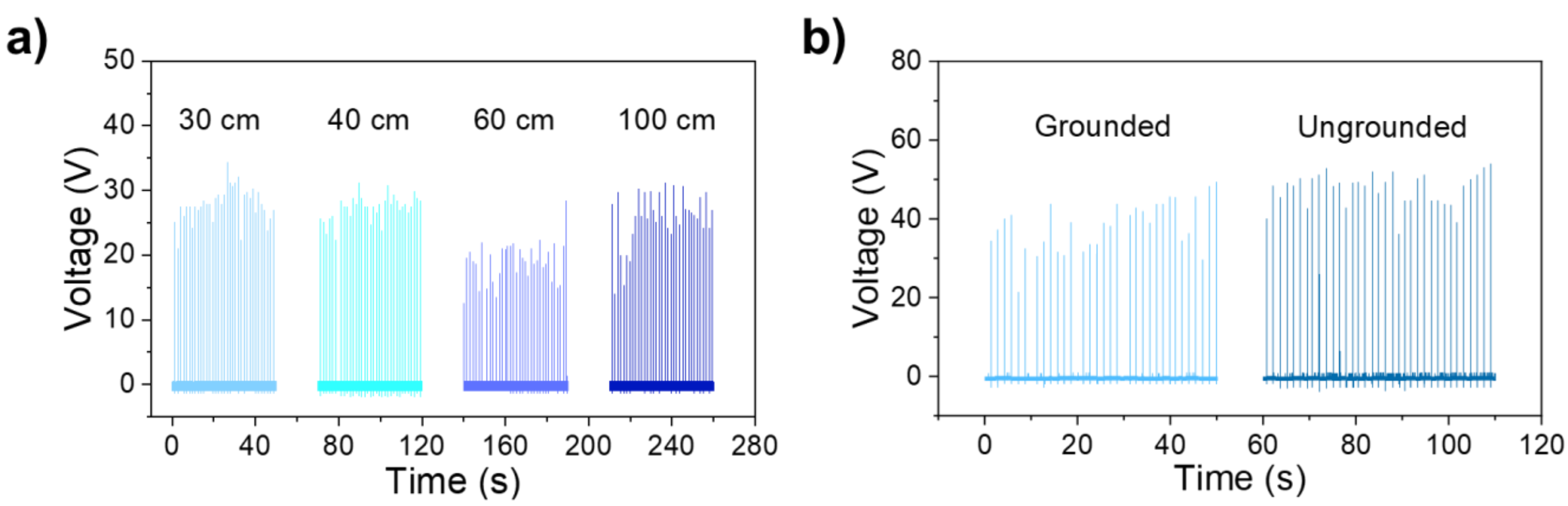

**Figure S10.** Open-circuit voltage peaks of the $CF_x$ D-TENG under rainwater droplets: a) effect of varying drop height from 30 cm to 100 cm, and b) influence of connecting or disconnecting the tip to ground.

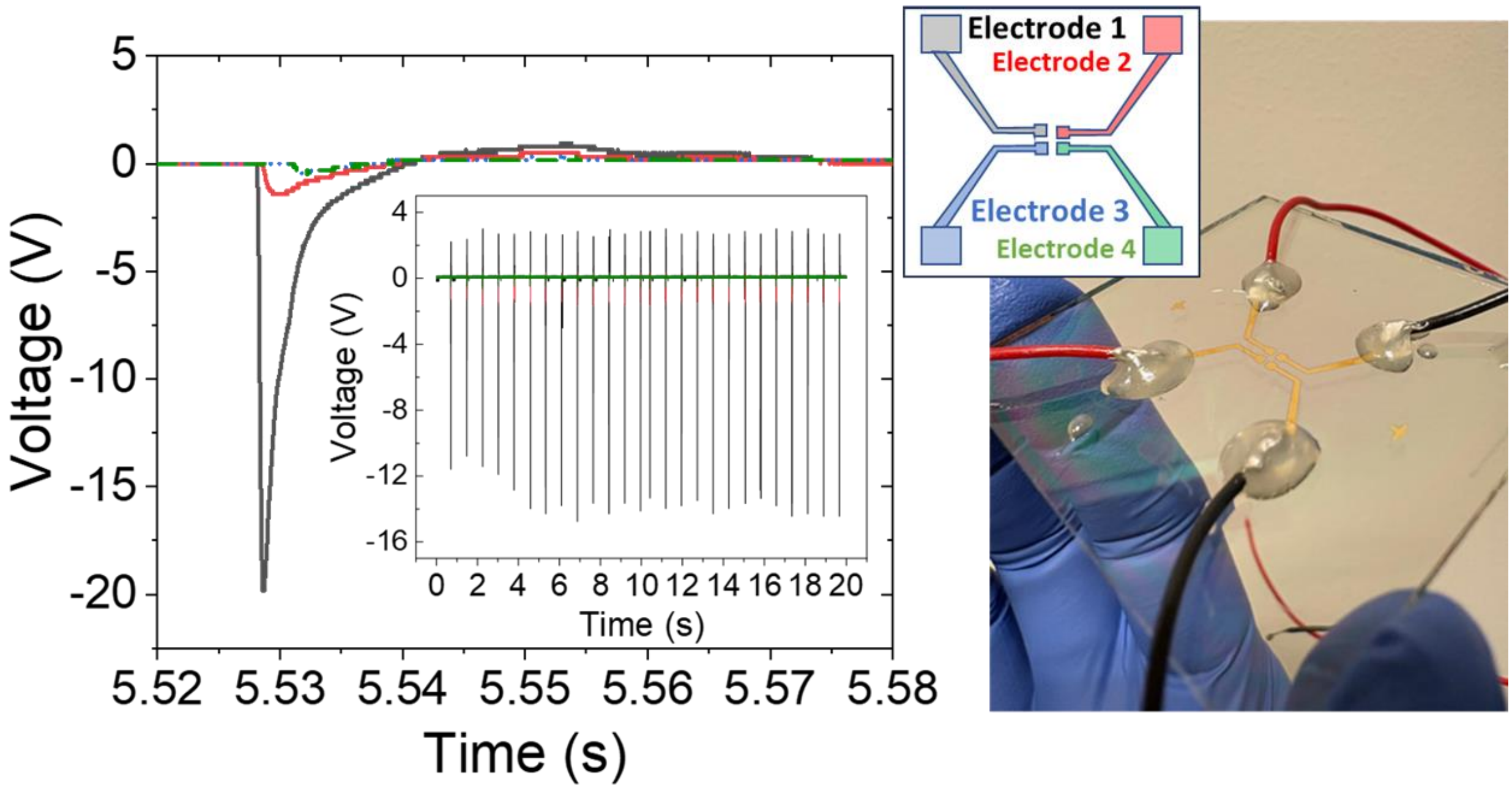

**Figure S11**. Analysis of open-circuit voltage for a 2×2 transparent $CF_x$ D-TENG array, including image and schematic of electrode distribution.

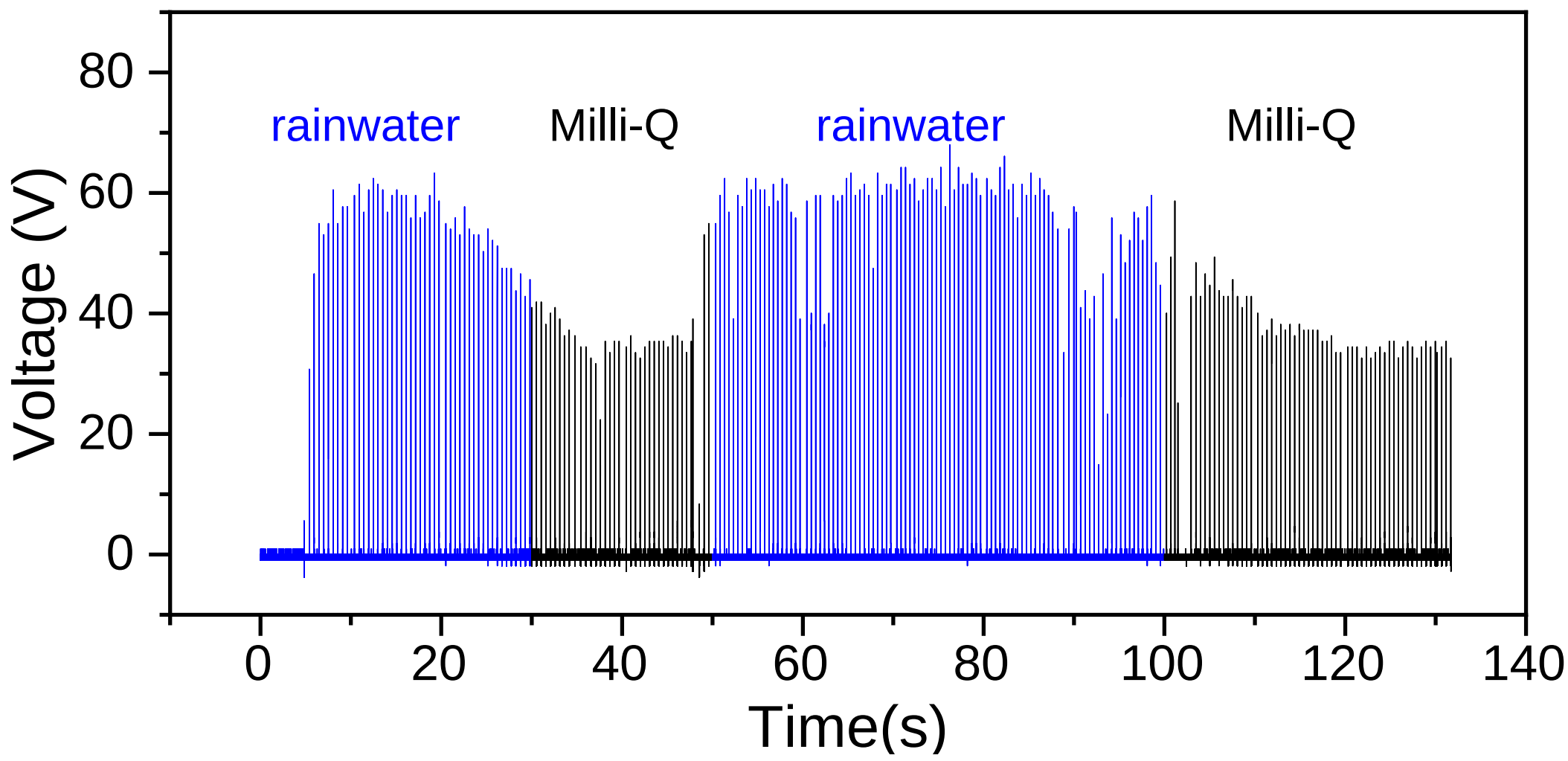

**Figure S12.** Open-circuit output voltage of the $CF_x$ D-TENG under rainwater/Milli-Q water cycling measured at standard conditions: 45º of tilting, 30 cm of height, ungrounded tip, and 3 Hz of frequency.

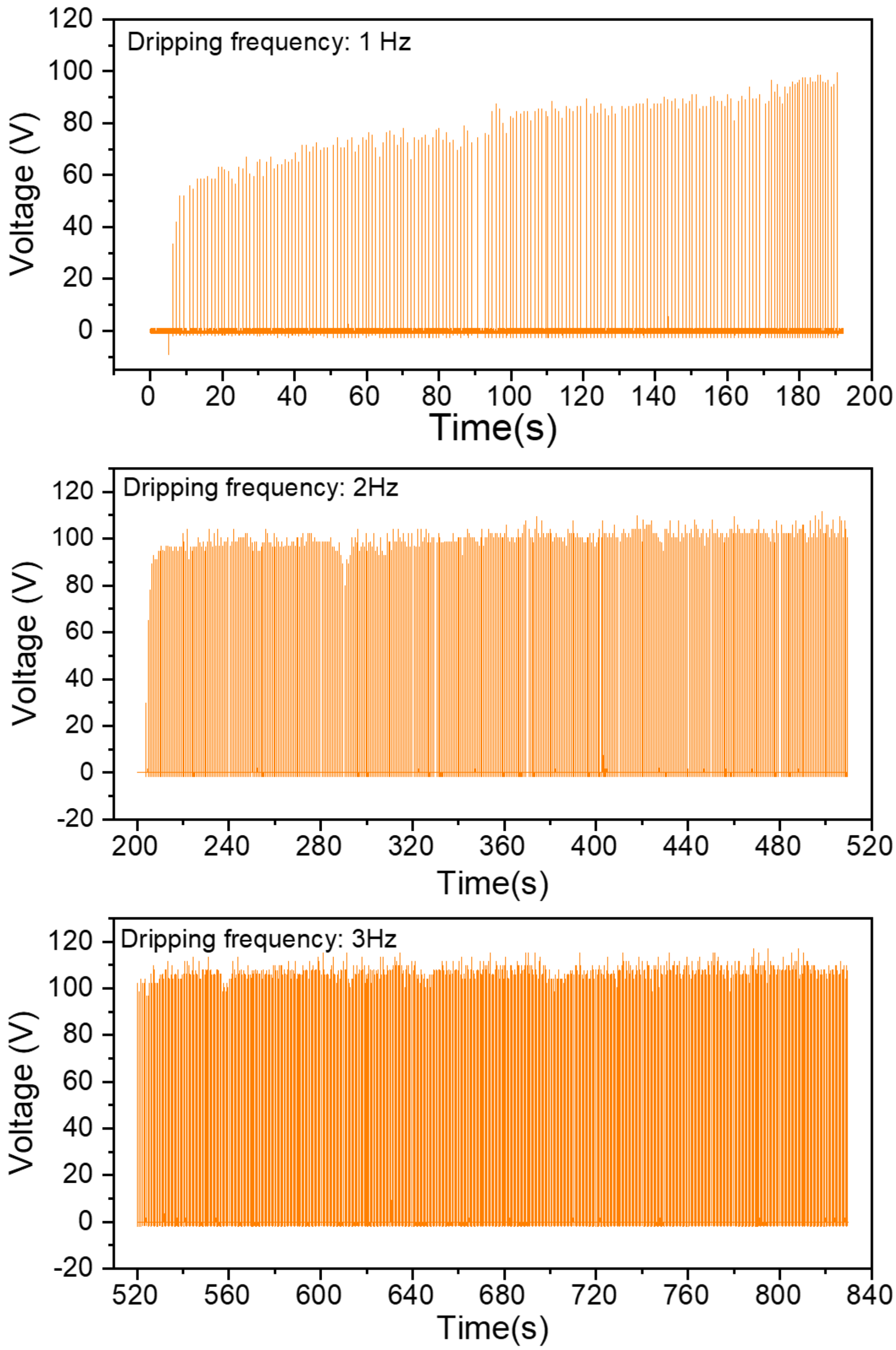

**Figure S13.** Charging phenomena of the $CF_x$ D-TENG device through the output voltage evolution with rainwater dripping increasing frequency from 1 Hz to 3 Hz, that reaches a plateau after 500 s.

**Table S3.** Overview of the state-of-the-art on hybrid solar and D-TENG harvesters.

| Solar technology | Nanogenerator | Cell efficiency | Nanogenerator output | Simultaneous energy harvesting (rain + solar) | Stability or reproducibility | Reference |
|---|---|---|---|---|---|---|
| c-Si | TENG (single electrode) | 15,71% 1 sun (AM 1.5 G, 100 mW $cm^{-2}$) | 12.18±1.75V 0.86±0.24μA 10.4μW (1.5MΩ) | No, capacitor charged curves | Not specified | [5] |
| c-Si | TENG (single electrode) | Not specified 1 sun (AM 1.5 G, 100 mW $cm^{-2}$) | 17.5V 3.5 mA·$m^{-2}$ 6mW·$m^{-2}$ (RL=5MΩ) | No, capacitor charged curves | Not specified | [6] |
| c-Si | TENG (single electrode) (3.2 cm x 4.2 cm) | 6% (0.7$cm^2$) (illumination level of 150 mW/$cm^2$) | 7V 128nA 0.265μW | YES (rain and light measuring) | Not specified | [7] |
| c-Si | TENG (single electrode) | 16% HC= 14% 1 sun (AM 1.5 G, 100 mW $cm^{-2}$) | 30V 4.2mA·m-$^2$ 170mW·$m^{-2}$ (RL=10MΩ) | YES (rain and light measuring) | Not specified | [8] |
| c-Si | TENG (array of single electrodes) | 19.61% 1 sun (AM 1.5 G, 100 mW $cm^{-2}$) | 6.95V 0.55μA 0.24mW·m-2 | YES (rain and light measuring) | YES (7 days) | [9] |
| c-Si | TENG (single electrode) | 12-13.6% HC= changes in 3% 1 sun (AM 1.5 G, 100 mW $cm^{-2}$) | 2.14V 33.0nA | No, capacitor charged curves | Not specified | [10] |
| c-Si | TENG (interdigitated single bottom electrode) | (4 V 30 mA, MiniSOLAR) 1 sun (AM 1.5 G, 100 mW $cm^{-2}$) HC= 1% of degradation in light transmittance | 18.4V 24.4μA | No, capacitor charged curves | Not specified | [11] |
| c-Si | TENG (top-bottom electrode) | 18.41% (3x3 cm) 1 sun (AM 1.5 G, 100 mW $cm^{-2}$) | 37.19V 7.59μA 147μW | No, capacitor charged curves | YES (7 days) | [12] |
| c-Si | TENG (single electrode) | 12.55% HC= 13.57% 1 sun (AM 1.5 G, 100 mW $cm^{-2}$) | 29.3V 32.0μA 453.4μW (RL=1MΩ) | No, capacitor charged curves | Not specified | [13] |
| c-Si | TENG (array of top-bottom electrode) | maximum power of 27.77 μW (power density is 37.03 mW $m^{-2}$ ) of solar cell at 500 Lux | 103V 46nC 80μA (3.2%) 48.89 W·$m^2$ (RL=2.5MΩ) | YES (rain and light measuring) | YES | [14] |
| c-Si | TENG (single electrode) 200μm | Not specified | 4.7V 0.35μA | No, capacitor charged curves | Not specified | [15] |
| c-Si | TENG (top-bottom electrode) AA=2.83$cm^2$ | 18.47% HC= 19.6% | 80.5V 21.5μA 1.27W·$m^2$ (RL=4.7MΩ) | No, capacitor charged curves | YES (7 days) | [16] |
| | | 1 sun (AM 1.5 G, 100 mW cm⁻²) | | | | |
| c-Si | TENG (single electrode) | 13.60%<br>HC= 14.6% | 23V<br>2.1µA<br>13.9µW | No, capacitor charged curves | Not specified | [17] |
| c-Si | TENG (interdigitated bottom electrode) 8x8cm | Not specified (1300 W m-² ) | 0.8µA, 6V, 0.34µW | No, capacitor charged curves | Not specified | [18] |
| c-Si | TENG (top-bottom electrode)<br>TL ~ 3-10 µm<br>AA= 1.2 x 1.2 cm² | 20.84%<br>1 sun (AM 1.5 G, 100 mW cm⁻²) | 80V, 110.5 mA/m², 1.06W/m² (RL=1MΩ) | YES (rain and light measuring) | YES (months) | [19] |
| Semitransparent polymer | TENG (top-bottom electrode)<br>TL ~50-160µm | 17.4 - 15.7%<br>HC= 10.1-8.4%<br>1 sun (AM 1.5 G, 100 mW cm⁻²) | ~100 V<br>~120 µA<br>~60 nC<br>2.62 W m⁻² | Yes (rain and light measuring) | YES (7 days) | [20] |
| Perovskite Solar Cell | TENG (top-bottom electrode)<br>TL ~1,3 µm<br>(2.5 × 2.5 cm² ) | 19.70%<br>HC=19.38%<br>1 sun (AM 1.5 G, 100 mW cm⁻²) | 110V<br>20µA<br>0.68mW | No, capacitor charged curves | YES | [21] |
| Perovskite Solar Cell | TENG (top-bottom electrode)<br>TL ~ 200nm<br>(2 x 2cm²) | 17.9 %<br>1 sun (AM 1.5 G, 100 mW cm⁻²)<br><br>Multisource = 15.7 %<br>0.5 sun (AM 1.5 G, 50 mW cm⁻²) | 110 V<br>0.6 mW,<br>40 W/m²<br>3µA/cm²<br>33.8 nJ ,<br>2.25 mJ/m²<br>($R_L$=476 kΩ) | YES (rain and drop measuring) + immersion measuring | YES (>5h) | This work |

**Supporting Information S14.**

*Perovskite Solar Cells fabrication.* FTOs were cleaned by brushing their surface with Hellmanex solution in water (2:98 vol %) and rinsed with deionized water. Then, substrates were sonicated in an ultrasonic bath for sequential 15 min cycles in Hellmanex solution, deionized water, acetone and iso-propanol. Finally, the FTO substrates were ozonized (UV/$O_3$ treatment) for 15 min in Ossila Ozone Cleaner. The Electron Transport Layer (ETL) consists of a 30 nm compact (c-$TiO_2$) layer deposited by the spray pyrolysis method (solution of 1 mL of titanium diisopropoxide bis(acetylacetonate) in 14 mL of absolute ethanol and sprayed on annealed glasses (450ºC for 30min) using oxygen as carrier gas). Before the fabrication of the next mesoporous (m-$TiO_2$) layer, the substrates were cooled down and treated with UV/$O_3$ again for 15 min. Then, 100 µL of a $TiO_2$ colloidal dispersion (adding 1 mL absolute ethanol to 150 mg of a commercial $TiO_2$ paste and stirring overnight) was spin-coated at 4000rpm for 10s followed by a temperature treatment up to 450ºC. After an additional UV/$O_3$ activation, the perovskite deposition was done over the FTO/c-$TiO_2$/m-$TiO_2$ inside a glovebox in a nitrogen atmosphere (both $O_2$ and $H_2O$ levels were under 0.5 ppm and the temperature below 28ºC). Perovskite solution (RbCsMAFA) was prepared by the mixture 1:1 M of $FAPbI_3$:$MAPbBr_3$ solutions (5:1%V, respectively) both dissolved in 1:4 %V DMSO:DMF. Then a 5%V of 1.7M CsI solution was added to DMSO and 5%V of RbI solution in 1:4 %V DMSO:DMF (0.2:99.8% mol, respectively). Two-step spin-coating processing has been applied: a slow step of 1000 rpm during 10 s followed by 6000 rpm during 20 s where 200 µL of chlorobenzene was added as antisolvent in the second 15 of the final spinning. Immediately, the samples were annealed on a hot plate at 100ºC for 60 min. As Hole Transport Layer (HTL), a solution of 70 mM Spiro-OMeTAD in chlorobenzene doped with LiTFSI (520 mg/mL in acetonitrile), FK209 in acetonitrile and 4-tert-Butylpyridine in a molar ratio of 0.5, 0.03 and 3.3, was filtered, and degassed. After that, a 100 µL of the previous Spiro-OMeTAD solution was spin-coated in two steps: 200 rpm and 6000 rpm for 10s and 10s, respectively. Finally, PSCs were taken out from the glovebox, approximately 0.5cm of these layers were removed from the upper part of the cell, and 80 nm of gold was deposited by thermal evaporation as a top electrode layer. Cells are left in under air inside a desiccator overnight. It should be noted that the results reported here were initiated about two years ago, using devices representative of our laboratory performance at that time (average PCE ≈ 13–14%). While these values are below current benchmarks, current higher-efficiency cells (also synthesized currently in our laboratory) generally exhibit greater intrinsic stability. Therefore, the benefits of $CF_x$-based encapsulation demonstrated here are expected to be even more pronounced in future work with state-of-the-art devices.

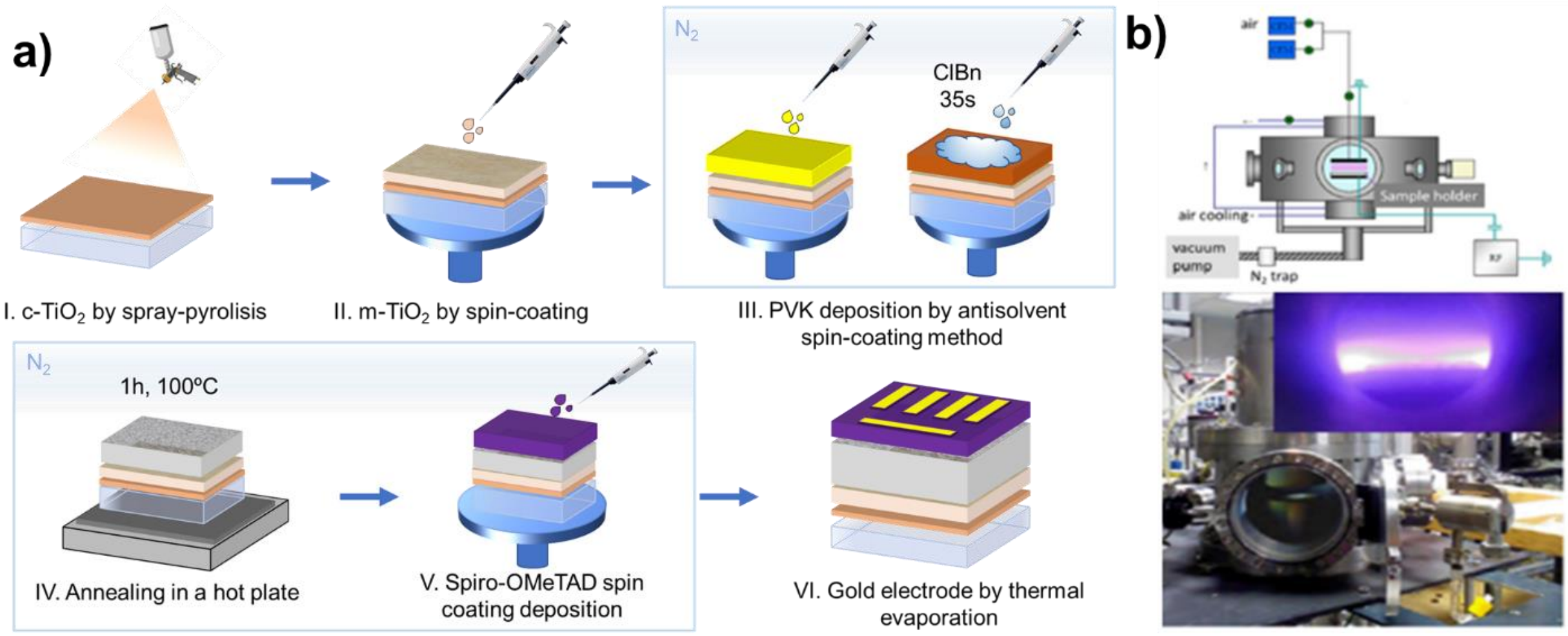

**Scheme S1.** Schematic of the different processes involved in the perovskite solar cells fabrication: a) sequential drawings corresponding to the steps performed for the perovskite solar cell implementation; b) diagram and picture of the RF plasma reactor used for the deposition of the water-repellent $CF_x$ coating.

**Supporting Information S15.** Estimation of PV Parameters for Non-Standard Conditions in Figure 6 and Figure 7. [1]

Several challenges affect the determination of PV parameters under these conditions:

1) **Lower Illumination Power:** I-V curves were measured at 50 mW/cm$^2$ (0.5 suns) by positioning the samples farther from the simulator to allow proper dripping on their surface.
2) **Impact of Wet Surface:** The wet surface altered the illumination, even on hydrophobic materials like $CF_x$. Dripping was tested before illumination to adjust positioning, meaning the surface was already wet when measurements began. This likely explains the initial rise in PCE during the first 30 minutes, as dripping combined with illumination led to temperature equilibration.

   This enhancement of the PCE can be linked to phenomena such as light soaking,[2] change in the refractive index during water dripping,[3] or other intrinsic stabilization mechanisms of the perovskite.[4]

3) **Difficulty in Determining Active Area:** Calculation of the active area, crucial for determining current density, could not be accurately determined. Illumination masks could not be used on the D-TENG side because water accumulated on mask edges, preventing proper water flow. Additionally, imprecise Au evaporation borders and, especially, the rough laser etching line (used to prevent electrode short circuits) complicated active area definition. Spin coating deposition on the rough etched borders further caused uneven layer growth (m-$TiO_2$, Perovskite, and Spiro-OMeTAD).

Taking into account all these considerations (especially 1 and 3 since the number 2 cannot be used for the calculation), the calculated PCE (Power Conversion Efficiency) of a solar cell is extracted from I-V curves as:

$$PCE = \frac{I_{SC} \cdot V_{OC} \cdot FF}{A \cdot P}$$

With *P= 50 mW/cm²*, *A = 0.36 cm²*, and the rest of the parameters ($I_{SC}$, $V_{OC}$, and FF) are extracted from the I-V curves. With these assumptions, an initial PCE of 15.7% was calculated (see Table 2).

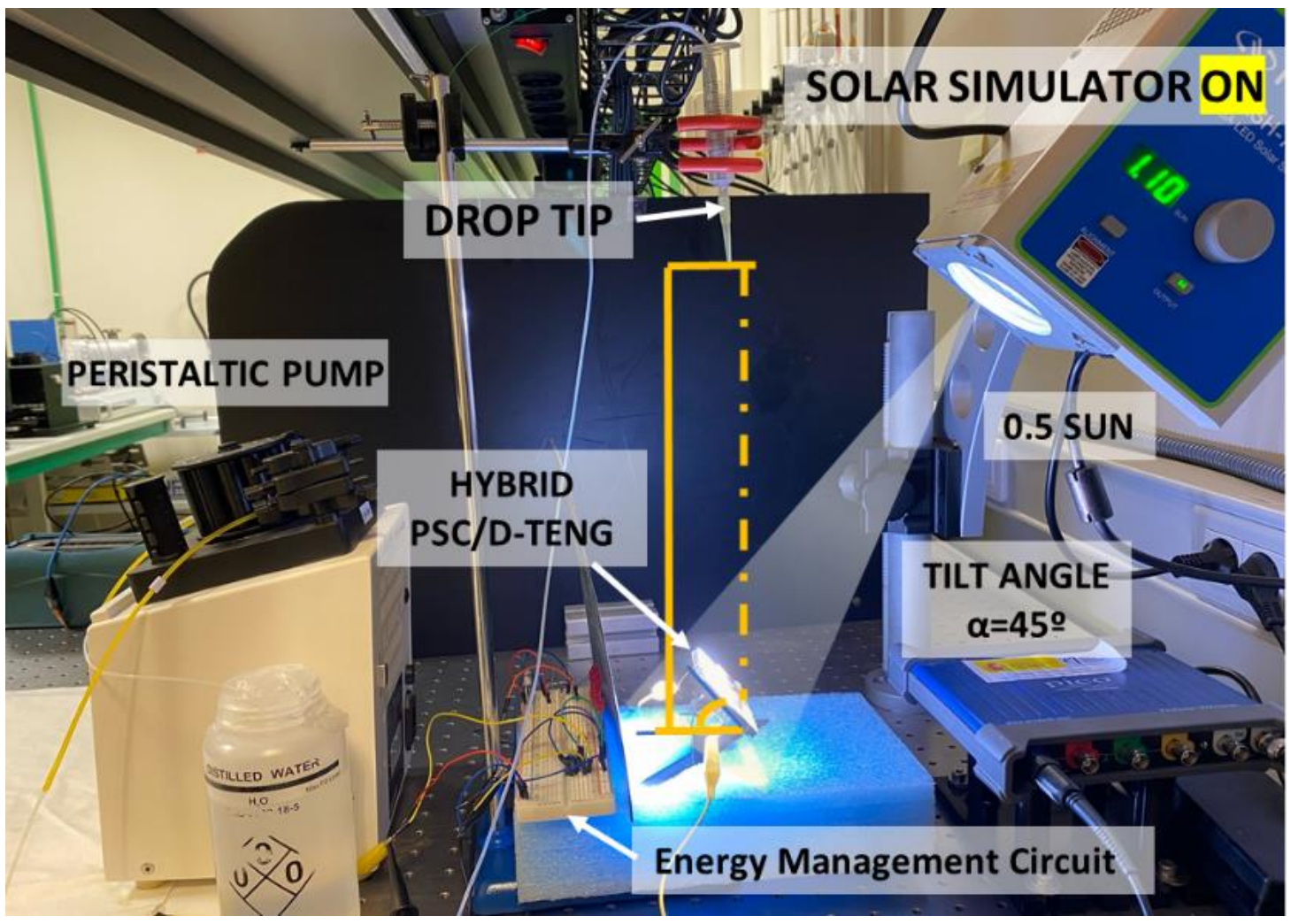

**Figure S14.** Photograph of the experimental setup for sequential and simultaneous drop-TENG and solar cell evaluation. The hybrid cells have been tilted 45º with respect to the vertical direction. The distance between the solar simulator and the cell and the illumination conditions in this configuration yield the equivalent of 0.5 suns.

Videos are available from the authors.